# Automated Logging Is Language-Sensitive: A Multilingual Benchmark and Empirical Study of LLMs

RENYI ZHONG, The Chinese University of Hong Kong, Hong Kong
YICHEN LI, The Chinese University of Hong Kong, Hong Kong
YULUN WU, The Chinese University of Hong Kong, Hong Kong
JINXI KUANG, The Chinese University of Hong Kong, Hong Kong
YINTONG HUO, Singapore Management University, Singapore
MICHAEL R. LYU, The Chinese University of Hong Kong, Hong Kong

Logging statements are central to debugging, failure diagnosis, and production observability, yet writing them requires developers to decide where to place a logging statement, which API and severity level to use, and what runtime information to expose. Automated logging aims to reduce this burden, but existing evidence remains dominated by Java-centric repository-snapshot datasets. Thus, Java-centered benchmarks do not show whether findings on model behavior and model selection generalize to other programming languages or to real-world maintenance changes that introduce logging statements. This paper presents MultiLogBench, a multilingual benchmark and empirical study spanning six programming language ecosystems. MultiLogBench contains 63,965 production-code repository-snapshot instances, 744 revision-history cases where developers introduce logging statements during maintenance, and a paired transformed revision-history branch for robustness analysis. Using seven contemporary large language models under a unified protocol, we evaluate logging-site localization, framework-anchor matching, severity prediction, message generation, variable recovery, and cascaded overall quality. Results show that automated logging is sensitive to language-ecosystem conditions. On repository-snapshot data, absolute performance levels and component profiles vary across language-specific subsets, with position prediction exhibiting the largest average within-model component gap. The six language-specific CCS leaderboards are nevertheless highly correlated, with a mean pairwise Spearman correlation of 0.912, and identify the same top two models; reordering is concentrated among the middle- and lower-tier models. Among the primary structural buckets, loop and nested-callable sites are the hardest. On revision-history data, the six-language mean declines across all displayed metrics, cross-language score dispersion persists, the component profile shifts toward greater framework-anchor dispersion, and the repository-snapshot top-two stability does not persist. Transformed inputs do not cause a broad same-direction performance collapse. These results show that single-language evaluation is informative for identifying the leading models in the controlled repository-snapshot benchmark, but is insufficient for estimating absolute performance across language ecosystems, diagnosing component-specific weaknesses beyond the evaluated language, or establishing a fine-grained model ordering below the top two that transfers across languages.

CCS Concepts: • **Software and its engineering** → *Maintaining Software.*

Additional Key Words and Phrases: Software Logging, Logging Statement, Logging Practice, Large Language Model, Multilingual Benchmark

Authors' Contact Information: Renyi Zhong, ryzhong22@cse.cuhk.edu.hk, The Chinese University of Hong Kong, Hong Kong; Yichen Li, ycli21@cse.cuhk.edu.hk, The Chinese University of Hong Kong, Hong Kong; Yulun Wu, ylwu24@cse.cuhk.edu.hk, The Chinese University of Hong Kong, Hong Kong; Jinxi Kuang, jxkuang22@cse.cuhk.edu.hk, The Chinese University of Hong Kong, Hong Kong; Yintong Huo, ythuo@smu.edu.sg, Singapore Management University, Singapore; Michael R. Lyu, lyu@cse.cuhk.edu.hk, The Chinese University of Hong Kong, Hong Kong.

## 1 Introduction

Logging statements serve as a critical mechanism for understanding runtime behavior in modern software systems. Developers depend on logs for failure diagnosis, execution path reconstruction, program state inspection, incident response, and production observability [23, 24]. Despite their importance, producing high-quality logging statements consistently presents significant challenges. Developers must make multiple interdependent decisions: determining insertion locations, selecting appropriate logging frameworks or APIs, assigning severity levels, crafting informative messages, and identifying relevant variables or expressions to capture. These decisions are inherently context-dependent, labor-intensive, and frequently governed by project-specific conventions [69]. This motivates substantial research interest in automated logging statement generation.

A growing body of work has investigated automated logging statement generation from multiple perspectives. Prior research has proposed techniques for generating complete logging statements [12, 13, 43, 72], improving logging text quality [73], and evaluating large language models (LLMs) for this task [74]. Together, these studies demonstrate substantial progress toward automating the task. However, the evidence used to assess that progress remains concentrated in narrow evaluation settings. Because logging-statement decisions depend on programming-language ecosystems and local development context, conclusions about model performance under one language and data setting may not transfer unchanged to others.

Two dimensions of transferability remain unresolved. First, current evidence is predominantly single-language, with a pronounced Java-centric bias. Representative studies, including LANCE [45], LEONID [44], SCLogger [34], Unilog [66], Fastlog [65], and PDLogger [15], have primarily trained on, evaluated against, or benchmarked using Java projects. Although recent work such as LogGen [32] has begun extending coverage beyond Java, the broader landscape remains limited. These language ecosystems differ in syntax, type systems, control-flow idioms, exception-handling mechanisms, namespace organization, logging framework designs, and project conventions. Evidence drawn mainly from one language ecosystem therefore leaves the cross-language transferability of model-performance and model-selection conclusions uncertain.

Second, most existing methods and datasets [34, 45] rely on *repository-snapshot data*. In such benchmarks, a developer-authored logging statement found in a fixed repository revision is treated as the gold target, and the model is evaluated by reconstructing that statement from its surrounding code context. This offline formulation is useful because it provides scalable, reproducible labels for controlled experimental comparison. However, it evaluates whether a model can reconstruct a logging decision already present in a completed code state, leaving open whether conclusions obtained in this setting persist when the target is anchored in an observed code-evolution event. Revision-history data extracted from authentic logging-statement-introducing commits provides a complementary maintenance-derived setting for examining this transfer [14]. Together with the predominantly Java-centered coverage of the closest snapshot studies [33, 59], this limitation leaves a two-dimensional evidence gap: which conclusions about model performance and model selection transfer across language ecosystems, and which persist for maintenance-derived logging-statement introductions.

To investigate this transferability, we construct MultiLogBench, a multilingual benchmark for automated logging statement generation spanning six language ecosystems: Java, Python, Go, C++, JavaScript, and C#. MultiLogBench combines three complementary data branches. The repository-snapshot branch provides the primary controlled benchmark for comparing the same models across the six language ecosystems. The revision-history branch uses authentic logging-statement-introducing revisions to test which repository-snapshot findings persist in a more maintenance-oriented setting. Following prior transformation-based robustness testing [33], the transformed revision-history branch complements these two branches by testing whether their aggregate findings remain stable under target-preserving surface variation.

Guided by this benchmark design, we study automated logging statement generation as a cross-language stability problem rather than merely a single-language performance optimization task. Our central question is which conclusions about absolute performance, component-specific behavior, and model ordering remain stable when the same models are evaluated across language-ecosystem conditions and in a more maintenance-oriented code-evolution setting. We organize the investigation around four research questions. **RQ1** examines how the evaluated models' performance and rankings vary across language-specific subsets of the shared repository-snapshot benchmark. **RQ2** investigates how the structural position of a logging statement within a function is associated with generation difficulty, distinguishing structure-linked variation from variation across language-specific subsets. **RQ3** assesses which aspects of the repository-snapshot multilingual findings persist when evaluation transitions to maintenance-oriented revision-history data. **RQ4** tests whether aggregate revision-history performance changes broadly under target-preserving surface transformations.

The repository-snapshot results show that model outcomes are sensitive to language-ecosystem conditions, but the implication depends on which conclusion is transferred. Absolute model performance and component profiles vary across language-specific subsets, and position prediction has the largest average within-model component gap. Conditional content outcomes also differ: for example, the JavaScript subset combines high framework-anchor and level accuracy after correct placement with lower position accuracy, whereas outcomes on C++ are below the multilingual mean across several content components. Among the primary structural buckets, loop and nested-callable positions are especially challenging because they require models to localize the intended logging statement within iterative updates or competing inner scopes. At the same time, the language-specific leaderboards exhibit high overall agreement: their mean pairwise Spearman correlation is 0.912, and Gemini-3.1-Pro and Claude-Sonnet-4.6 remain first and second across all six subsets. Single-language leaderboards are therefore informative for identifying the leading models in this controlled benchmark, but they do not reliably transfer absolute performance levels, component-specific weaknesses, or fine-grained ordering below the top two.

The validation branches further show that the repository-snapshot findings do not transfer uniformly to maintenance-derived conditions. On revision-history data, the six-language mean declines across all displayed metrics, the component profile shifts toward greater framework-anchor dispersion, and cross-language score dispersion persists, whereas the repository-snapshot top-two stability does not. Under target-preserving surface transformations, however, aggregate revision-history performance does not exhibit a broad same-direction collapse.

This paper makes the following contributions:

- We construct MultiLogBench, a multilingual benchmark for automated logging statement generation across six language ecosystems. Unlike benchmark designs that rely only on frozen repository snapshots, MultiLogBench integrates a repository-snapshot core branch with two evaluation-only validation branches: revision-history data mined from authentic revisions in which developers introduce logging statements and a paired transformed revision-history branch for robustness analysis.

- We formulate automated logging statement generation as a cross-language stability problem and provide a unified evaluation protocol for studying it. This protocol evaluates the task across languages, structural positions, models, and output components, allowing the study to distinguish overall generation quality from specific failure sources such as logging-site localization, framework-anchor matching, severity prediction, message construction, and variable recovery.
- We conduct a controlled multilingual comparison that establishes an empirical basis for assessing the transferability of model-performance and model-selection conclusions across language ecosystems. The resulting evidence characterizes transferability at the levels of absolute performance, output components, structural contexts, and model ordering without assuming that conclusions from any single language-specific subset generalize unchanged.
- We complement the repository-snapshot comparison with maintenance-oriented validation on authentic logging-statement-introducing revisions and a paired transformation-based sensitivity analysis. Together, these branches provide evidence for assessing whether snapshot-derived conclusions persist under observed code-evolution events and target-preserving surface variation.

To support reproducibility and future research, we publicly release the source code and benchmark data of MultiLogBench at https://github.com/logresearch/MultiLogBench.

The remainder of this paper is organized as follows. Section 2 reviews the background and related work on automated logging and empirical studies of logging practice. Section 3 introduces the construction of MultiLogBench. Section 4 describes the experimental design. Section 5 presents the experimental results for RQ1–RQ4 and a qualitative analysis of representative failure patterns. Section 6 discusses lessons learned, threats to validity, and future work. Section 7 concludes the paper.

## 2 Background and Related Work

### 2.1 Automated Logging

Research on automated logging has progressed from developer decision support to end-to-end generation systems. Early studies primarily focused on helping developers decide *whether*, *where*, and *what* to log. For example, prior work proposed learning-based support for logging decisions [76], investigated log-level recommendation from code context [27, 30], and explored level selection based on developer intention or verbosity preferences [3, 42, 60]. Complementary work examined where to place logging statements, including recommendation within code blocks [39], industrial guidance on log placement [5], and overhead-aware optimal placement [69]. Together, these studies established the main subproblems of automated logging, but most stopped short of generating complete logging statements.

Subsequent work moved toward generating logging-statement content directly from source code. Representative approaches include clone-based prediction [18], neural machine translation for logging text generation [12, 13], and deep-learning methods for generating complete logging statements [44, 45]. More recent systems extended the task to end-to-end generation and insertion, such as Unilog [66], Fastlog [65], SCLogger [34], PDLogger [15], and Autologger [72]. The most recent line of work increasingly leverages large language models, including empirical studies of overall effectiveness [33] and model-scale trade-offs [74]. Overall, the literature reflects a clear shift from recommendation-oriented support to fully automated logging generation. At the same time, robustness, benchmark coverage, and evaluation design remain active research challenges, as highlighted by AL-Bench [59] and recent runtime-feedback-oriented work [62]. More specifically, Li et al. [33] examined eleven LLMs for Java logging-statement content at specified logging points using LogBench-O and the transformed LogBench-T, whereas AL-Bench [59] standardized Java snapshot evaluation and added compilability and runtime-log measures.

This progression also shifts the central evaluation question. Once automated logging is framed as a full generation task rather than merely a set of local recommendation subproblems, the key issue is not only whether models can generate logging statements, but also whether empirical findings about their performance generalize across language ecosystems and evaluation settings.

### 2.2 Empirical Studies on Logging Practice

Empirical studies on logging practice have examined the topic in both industrial settings [17] and open-source systems [57], showing that developers' logging decisions are non-trivial and closely tied to program comprehension, diagnosis, and maintenance [21]. Subsequent work linked logging characteristics to code quality [56], examined logging-library migrations [25], characterized logging use in Java open-source projects [7], and studied the long-term stability of logging statements [26]. Taken together, these studies indicate that logging statements are program artifacts relevant to development and maintenance, shaped by project conventions, diagnostic goals, and software evolution rather than incidental code fragments.

Recent empirical studies expanded this line of inquiry from prevalence to quality and risk. Prior work reported logging anti-patterns [6], log-related issues [22], duplicate logging smells [37, 38], logging-code-issue-introducing changes [8], configuration practices [70], logging statement readability expectations [36], and defect repair for existing logging statements [63, 73], showing that logging-quality problems are widespread and evolve alongside the software itself. The literature also reports scenario-specific findings on mobile apps [67], Java logging utilities [9], test code [68], privacy leakage [1, 75], sensitive-information exposure [71], exception stack-trace logging [31], and machine-learning-based applications [16]. Complementary qualitative and exploratory studies further examine the benefits and costs of logging [29], logging practice in open-source software [53], and whether industrial logs capture developers' intended information [54]. Taken together, these studies show that logging practice is strongly context dependent and varies across projects, application scenarios, and technical settings.

A separate line of work has evaluated automated logging methods and benchmarks. However, existing evaluations, including deep-learning generation [45], contextualized generation [34], small-LLM evaluation [74], and hybrid code-analysis-plus-LLM generation [32], have largely remained limited to single-language settings or repository-snapshot data. Considered together, these two literature streams leave two unresolved methodological questions: whether empirical conclusions about automated logging generalize across language ecosystems, and whether they retain external validity when evaluation moves from repository-snapshot data toward authentic logging-statement-introducing changes.

Our work addresses this combined gap. More precisely, we use *cross-language stability* to denote the stability of empirical findings under a shared multilingual protocol rather than a standalone performance construct. We compare model outcomes across six language-ecosystem conditions, defined here as programming-language settings together with their logging APIs and conventions, and decompose performance into position, framework, level, message, and variable dimensions. We further test which aspects of the multilingual model-performance findings persist when evaluation moves from repository-snapshot data to revision-history data mined from authentic logging-statement-introducing changes, followed by a separate robustness analysis using data produced by target-preserving surface transformations. In this sense, our study complements prior empirical research on logging practice: earlier work documents how logging statements are used and where quality problems arise, whereas we examine whether conclusions about model performance remain stable across ecosystem conditions. This motivation leads directly to the benchmark construction choices described in the next section.

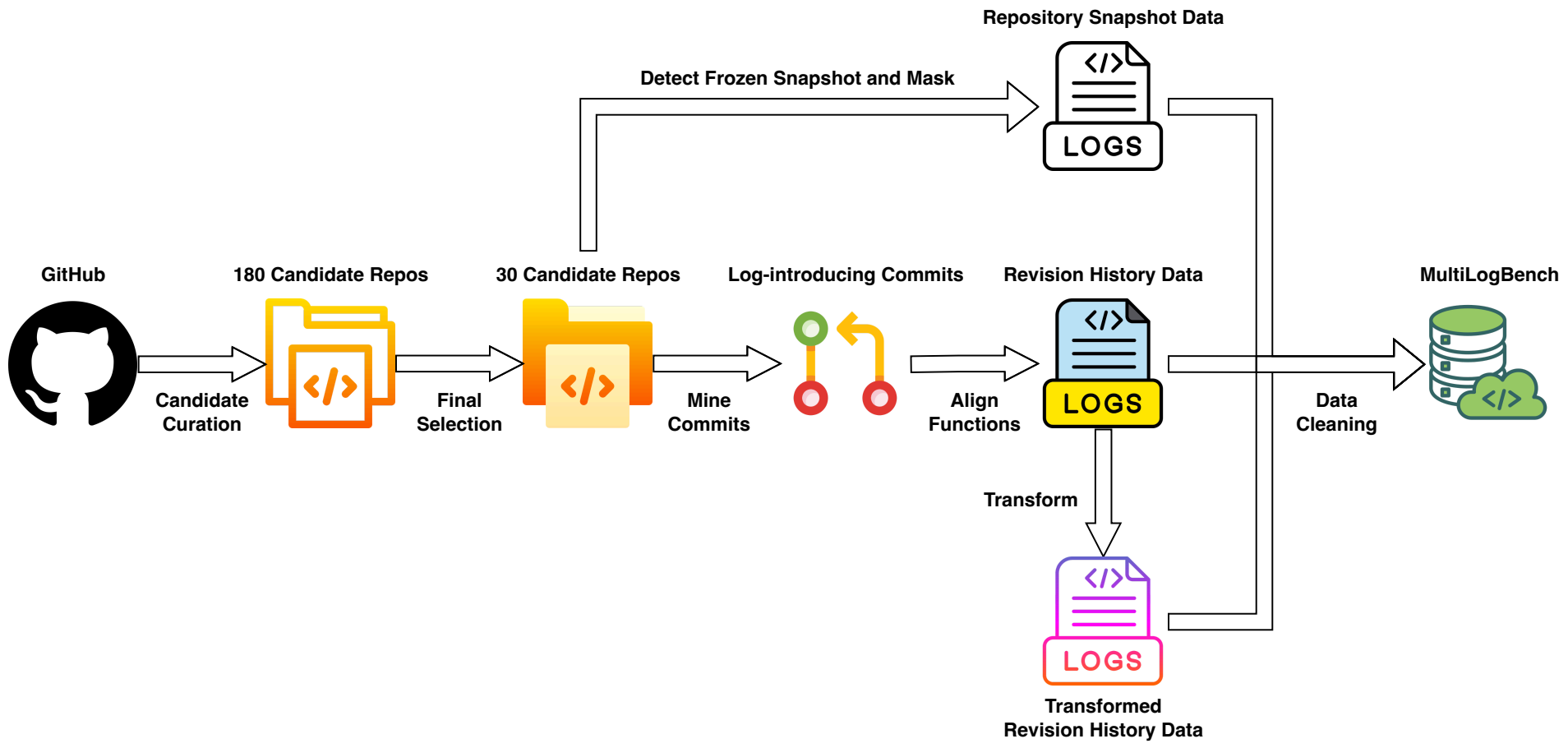

Fig. 1. Overview of the benchmark construction pipeline.

## 3 Benchmark Construction

This section describes the construction of the benchmark. We first define the data types and repository pool, and then present the mining procedures, unified postprocessing, transformed-data construction, and final split statistics. Figure 1 provides an overview of the benchmark construction pipeline.

### 3.1 Data Types and Sample Definition

We construct two data types: **repository-snapshot data** and **revision-history data**. Repository-snapshot data serves as the primary evaluation source. Its instances are mined from a frozen snapshot of each selected repository, where each target is a developer-authored logging statement that already exists in that revision. The statement is removed from its surrounding callable to form the model input and is retained as the gold output. To the best of our knowledge, prior studies of automated logging generation have primarily relied on this construction [15, 34, 65, 66], largely because it can be mined at scale. Figure 2 summarizes this target-provenance contrast and the shared prediction format of the two data types.

Revision-history data is instead constructed from logging-statement-introducing changes in repository histories. Each instance is anchored at a commit that adds a logging statement to a named method. After filtering, the aligned callable-level edit contains no substantive non-logging changes. The model receives only the pre-change parent callable, while the logging statement introduced in the child revision serves as the gold output.

Revision-history data more closely reflects logging decisions made during software evolution and therefore provides a more ecologically valid complement to repository-snapshot evaluation [14, 73]. It allows us to examine whether findings obtained in the controlled snapshot setting also generalize to code-evolution contexts. However, such instances require commit-level reconstruction, callable alignment, and strict filtering, making them substantially more difficult to obtain. We therefore use repository-snapshot data as the primary evaluation source and revision-history data as a smaller complementary validation set.

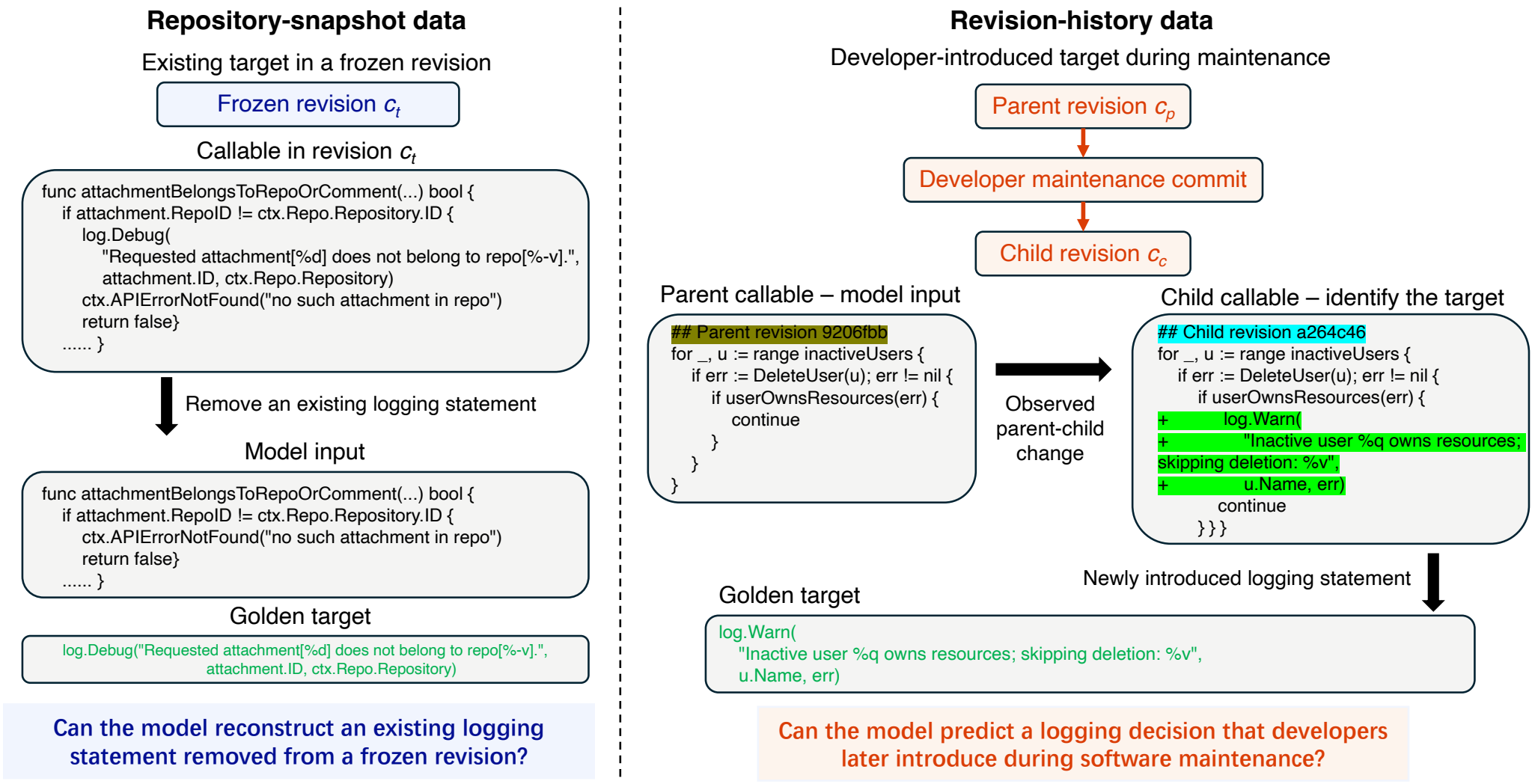

Fig. 2. Contrasting the construction and target provenance of repository-snapshot and revision-history data.

### 3.2 Language Scope and Repository Selection

We target six language ecosystems: Java, Python, Go, C++, JavaScript, and C#. This scope is broad enough to capture meaningful cross-language variation in logging practice while remaining small enough to support controlled benchmark construction. In practical terms, these ecosystems remain among the most widely used in developer communities [19, 58]. Methodologically, they span distinct programming and logging styles, including object-oriented enterprise systems (Java and C#), dynamic and scripting-heavy software (Python), cloud-native and structured-logging settings (Go), systems software with macro- or stream-based logging (C++), and the JavaScript web and toolchain ecosystem.

We designed the repository selection procedure to balance ecosystem coverage and benchmark controllability. We first queried GitHub for repositories [10] in each target language and retained those satisfying the following criteria:

- at least 200 commits, to filter out short-lived or immature projects.
- at least 5 contributors, to prefer repositories with collaborative rather than individual-only development.
- at least 1,000 stars, to bias the pool toward widely used and visible projects.
- at least 10,000 lines of code, to exclude trivial codebases unlikely to exhibit stable logging practices.
- created on or before January 1, 2020, to retain projects with sufficient development history.
- last commit on or after May 1, 2024, to ensure that the selected repositories still reflect current development practice.
- not being a fork, to avoid duplicated development histories and near-redundant project variants.
- an open-source license, to ensure that the selected repositories can be used and redistributed as a research artifact.

Table 1. Selected repositories and per-repository average statistics by language.

| Lang. | Repositories | KLOC | SrcF | LogF | Prod. Inst. | LogF% |
|---|---|---|---|---|---|---|
| Java | kafka, dbeaver, spring-framework, keycloak, dubbo, JeecgBoot, openapi-generator, nacos, canal, netty | 882.0 | 5533.6 | 428.8 | 1679.4 | 11.5 |
| Python | airflow, ray, faceswap, transformers, home-assistant-core, freqtrade, geekcomputers-python, jax, detectron2, scrapy | 818.4 | 3626.5 | 306.0 | 1565.8 | 15.6 |
| Go | gitea, tidb, kubernetes, go-ethereum, golang-go, traefik, trivy, prometheus, syncthing, gogs | 949.7 | 3902.3 | 229.3 | 1492.1 | 13.1 |
| C++ | nebula, grpc, brpc, typesense, mediapipe, folly, foundationdb, tesseract, opencv, duckdb | 651.5 | 2307.8 | 167.6 | 703.9 | 12.1 |
| JavaScript | react, phaser, three.js, webpack, parcel, create-react-app, prettier, ghost, pdf.js, anime | 532.0 | 2843.6 | 147.7 | 357.9 | 8.1 |
| C# | jellyfin, sonarr, radarr, aspnetcore, abp, playnite, runtime, jackett, roslyn, files | 1706.8 | 7253.3 | 146.8 | 597.4 | 6.9 |

*Abbreviations:* KLOC = thousand lines of code; SrcF = scanned source files; LogF = source files containing logging statements; Prod. Inst. = production-code repository-snapshot instances retained after source-role filtering; LogF% = percentage of logging files among source files. All numeric columns report unweighted averages over the ten selected repositories in each language.

These criteria remove short-lived, inactive, trivial, or legally ambiguous projects before any logging-specific inspection, making the repository pool more comparable across language ecosystems.

We then refined the candidate pool for each language in two passes. First, we ranked the surviving repositories by star count, removed obviously non-representative projects such as tutorials, demos, and templates, and retained the top 30 repositories per language as a candidate pool for logging inspection. Second, because downstream instance construction requires a sufficient number of logging cases per language, we selected 10 repositories per language by prioritizing logging richness, measured by the number of identifiable logging statements and the number of source files containing logging statements. This second pass yields a benchmark-oriented sample rather than a population-representative census, while preserving the project maturity and ecosystem relevance enforced in the first pass. The resulting 60 repositories provide sufficient logging density for downstream instance construction without sacrificing the cross-language comparability established by the earlier screening criteria. Table 1 lists the final repositories together with language-level average statistics.

### 3.3 Data Mining Procedure

*3.3.1 Mining Repository-Snapshot Data.* We construct repository-snapshot data from frozen revisions of the selected repositories using a two-stage mining protocol. The first stage performs lightweight file-level screening with language-specific patterns for common logging APIs. This stage is intentionally permissive: its purpose is not to define benchmark instances, but to narrow the search space to files likely to contain developer-authored logging statements. The second stage applies syntax-aware analysis using tree-sitter parsers [4] and language-specific logging rules to detect logging invocations at the statement level. A purely textual search would scale easily but produce many false positives; the two-stage design avoids this problem by restricting detailed parsing to a pre-screened candidate set.

After identifying a candidate logging statement, we attribute it to the nearest enclosing named function or method and construct the instance around that callable. This step is central to the benchmark design because the callable provides the unit of context that is most naturally comparable across languages. Statements that cannot be mapped to a named callable are excluded rather than heuristically attached to anonymous callbacks, initialization blocks, or other ambiguous scopes. When a callable contains multiple eligible logging statements, we create one instance per target statement by removing only that statement and leaving the remainder of the callable unchanged. For class member methods, we retain a minimal enclosing wrapper when necessary so that the extracted context remains syntactically complete and readable.

The mining protocol further incorporates filtering and normalization decisions intended to support scientific comparability. We exclude generated files, vendored code, and third-party dependencies because the goal is to capture developer-authored logging behavior rather than inherited artifacts. At the same time, we retain non-production files such as tests, examples, and documentation, and explicitly annotate their source roles so that later analyses can derive stricter subsets without altering the construction procedure.

*3.3.2 Mining Revision-History Data.* We construct revision-history data by mining the version histories of the selected repositories for logging-statement-introducing changes. Unlike repository-snapshot data, which is defined over a frozen revision, this variant is intended to approximate the maintenance setting in which developers decide to add a logging statement while evolving existing code [73]. The central methodological challenge is therefore not merely to find commits that contain logging statements, but to isolate cases in which a newly added logging statement can reasonably be operationalized as the primary local logging decision of an edit rather than as a side effect of broader refactoring, file movement, or batch maintenance activity.

To achieve this goal, we use a commit-level mining procedure that begins with the revision history of each selected repository and applies strict filtering before instance construction. We consider only non-revert commits whose diffs are limited to modified source files in the target language, excluding renames, copies, and binary changes. Within the surviving commits, we compare the parent and child revisions of each changed file, parse both versions, and identify logging statements that appear in the child revision and overlap with added diff lines. We further restrict the candidate set to production code so that this smaller, evaluation-only branch remains aligned with the production-code focus of the final benchmark splits. We also retain only commits with a small number of added logging statements, because commits that introduce many logging statements at once are more likely to reflect a broad instrumentation effort than a localized logging decision.

Instance construction then proceeds through callable alignment across revisions. Each added logging statement is first mapped to its enclosing named callable in the child revision, and we then attempt to locate the corresponding callable uniquely in the parent revision. A candidate is rejected if the added logging statement cannot be attributed to a named callable, if the parent callable cannot be identified unambiguously, or if an equivalent logging statement already exists in the parent version. We then remove all added logging statements from the child callable and compare the resulting code with the parent callable to estimate non-logging change. Only cases with no remaining substantive non-logging edits are retained. This restriction does not establish developer intent, but it provides a conservative operationalization of a logging-statement-introducing decision by filtering out cases in which the added logging statement is tightly entangled with simultaneous behavioral modification.

For each accepted case, the model input is the parent-version callable, and the gold output is the logging statement added in the child revision; we additionally retain the child-version callable and associated commit metadata for traceability. The resulting revision-history data is smaller and more selective than repository-snapshot data, but it offers a complementary view of automated logging generation under realistic code evolution while maintaining a sufficiently controlled sample boundary for cross-language comparison.

### 3.4 Postprocessing and Cleaning

After raw mining, we apply a unified postprocessing pass to all mined instances before any downstream analysis. This step serves two purposes: removing minor extraction artifacts introduced by automated instance construction and adding a normalized target representation that improves cross-language comparability by reducing sensitivity to language-specific surface variation. On the artifact side, we repair formatting gaps introduced when the gold logging statement is removed from the enclosing callable. On the normalization side, we map log levels to a shared severity label set, deduplicate and standardize variable lists, and derive a normalized logging statement that separates the message template from its variable bindings together with a style annotation. Both dataset splitting and transformed-dataset construction operate on these cleaned corpora rather than on the raw mining outputs.

### 3.5 Constructing the Transformed Dataset

To assess robustness to possible pretraining overlap and memorization, we derive a transformed dataset from the cleaned revision-history data. Both modern code LLMs and general-purpose LLMs are trained on large-scale public code corpora, creating a meaningful risk that benchmark instances mined directly from GitHub may overlap with the models' pretraining data. Prior studies have documented overlap between code-generation benchmarks and open training corpora, higher model performance on contaminated subsets, and multiple leakage pathways in code-generation evaluation [46, 51]. In this setting, strong performance on the original benchmark may reflect a mixture of genuine inference and memorization [55, 64]. Following prior work on contamination-aware evaluation in automated logging [33], we use a paired transformation design that reduces exact surface-form overlap while preserving the target logging decision. This design enables a sensitivity analysis of aggregate performance under the applied rewrites.

The transformed dataset is built by applying conservative source rewrites to each accepted instance with the aim of preserving the benchmark task definition [33]. Starting from the original callable, we first rematch the target logging statement in the parsed ASTs and identify protected regions that should remain untouched, including the target logging statement itself, its string literals, recent variable definitions relevant to the statement, and the control conditions governing the logging site. We then generate candidate rewrites from a fixed family of readability-preserving transformation rules, as shown in Table 2. These transformations are designed to alter the surface realization of the code without changing the logging target or removing the contextual signals on which developers rely when composing logging statements. We therefore treat this pipeline as preserving task-relevant local evidence for evaluation rather than as establishing full program equivalence of the transformed callable.

Each transformed candidate is validated before being admitted to the dataset. The transformed callable must remain parseable, the target logging statement must be rematchable after transformation, and its statement text, severity level, message payload, variable set, and normalized target representation must all remain unchanged. We additionally require that the transformed code introduce no new identifiers, thereby reducing the chance that the transformation itself injects new semantic evidence. These checks establish target-level equivalence for evaluation rather than formal equivalence of complete program behavior. When multiple valid rewrites are available for the same instance, we select one deterministically using a stable hash-based rule, which makes the transformed dataset reproducible while avoiding ad hoc manual choice. When no valid rewrite exists, we retain the original callable so that the transformed branch remains one-to-one paired with the historical pool. The final transformed dataset therefore preserves the original logging task definition while reducing exact text overlap wherever rewriting succeeds, providing a principled basis for analyzing memorization sensitivity and robustness beyond the original mined benchmark.

Table 2. Implemented code transformation strategies for constructing surface-distinct inputs while preserving the benchmark target definition. Red-highlighted code denotes the original form, and green-highlighted code denotes the transformed form.

| Transformation type | Description | Example |
|---|---|---|
| Parenthesis Insertion | Inserts redundant parentheses into selected expressions in syntax patterns where local behavior is intended to remain unchanged. | `result = a + b * c`<br>`result = (a + b * c)` |
| Constant / Literal Surface Rewrite | Rewrites the surface form of non-logging numeric literals while preserving their value. | `size = 1024`<br>`size = 0x400` |
| Condition Neutral Rewrite | Applies conservative syntactic rewrites to conditions while intending to preserve their local truth conditions. | `if ready:`<br>`if (ready) and True:` |
| Redundant Block / Suite Normalization | Expands single-line control-flow bodies into explicit block/suite form without changing statement order or scope. | `if ok: return value`<br>`if ok:`<br>`    return value` |
| Comment / Docstring Stripping | Removes comments and selected docstring-like text unrelated to the target logging statement; this is treated as task-preserving rather than as a guarantee of full program equivalence. | `x = 1 # temp fix`<br>`x = 1` |
| Else-if / Nested-if Normalization | Rewrites an `elif` branch into a nested `else: if` form, or vice versa, in patterns intended to preserve branch behavior. | `elif failed: handle()`<br>`else:`<br>`    if failed: handle()` |
| Ternary ↔ If/Else Normalization | Rewrites a simple `if/else` assignment or return into a conditional expression, or vice versa, in patterns intended to preserve the local result. | `if flag: x = 1`<br>`else: x = 0`<br>`x = 1 if flag else 0` |

### 3.6 Dataset Split and Statistics Summary

We derive the train, validation, and test splits exclusively from the repository-snapshot branch, as this is the only branch intended to support future model training and hyperparameter tuning. Starting from the cleaned snapshot corpus, we retain only production-code instances and assign all records from the same language, repository, and function-position group to a single split so that no group is divided across splits. We then apply deterministic repository-internal stratified splitting toward an 80/10/10 ratio while jointly balancing split size and log-level distribution. Repositories with fewer than ten eligible groups, or those that cannot satisfy the ratio constraint under group-level assignment, are allocated entirely to training. This procedure yields 51,120 training instances, 6,426 validation instances, and 6,419 test instances from 63,965 production-code instances in total. The multilingual composition remains highly consistent across splits: the maximum deviation in language share between any split and the full production corpus is 0.23 percentage points.

The revision-history and transformed-revision-history branches are retained as evaluation-only resources. Across the 60 full-history repositories used for revision-history mining, the recorded code changes contain 1,645,840 commits; after initial candidate filtering for added logging statements, 13,095 commits remain, and the subsequent callable-level alignment and cleaning steps retain 744 cases from 41 repositories in which a logging statement is newly added. Our default setting retains commits with at most 150 added-or-deleted source lines and accepts an aligned callable only when, after masking newly added logging statements, it has zero remaining non-logging edits relative to the parent callable. Table 3 reports the per-language sample counts and, in parentheses, the corresponding numbers of repositories contributing at least one case for this accepted pool and each construction-level sensitivity variant. We use this default setting because it preserves a clean local evaluation condition: moderately tightening the changed-line threshold leaves the pool largely stable, whereas relaxing the non-logging-edit constraint substantially changes the benchmark size and language composition. The transformed evaluation set is constructed as a one-to-one paired counterpart of the revision-history data. For each accepted case, the surrounding callable is rewritten using the target-preserving surface transformations described above whenever a valid rewrite is available; the gold logging statement, severity level, message payload, and variable set are preserved exactly. Of the 744 paired cases, 625 (84.0%) undergo a genuine rewrite, whereas 119 (16.0%) retain their original form because no applicable transformation is found.

Table 3. Revision-history sample and repository counts by language under construction-level sensitivity configurations. Each numeric cell reports the number of samples, with the number of contributing repositories in parentheses.

| Configuration | Java | Python | Go | C++ | JavaScript | C# | Total |
|---|---|---|---|---|---|---|---|
| Default Setting | 99 (10) | 426 (8) | 108 (7) | 23 (6) | 54 (7) | 34 (3) | 744 (41) |
| Changed lines ≤ 50 | 94 (10) | 418 (8) | 105 (7) | 22 (5) | 43 (6) | 31 (3) | 713 (39) |
| Single changed file + single added logging statement | 34 (9) | 190 (7) | 33 (7) | 8 (4) | 14 (5) | 4 (2) | 283 (34) |
| Allow ≤ 1 remaining non-logging edit | 516 (10) | 1,074 (9) | 531 (10) | 64 (8) | 149 (8) | 84 (3) | 2,418 (48) |
| Allow ≤ 8 remaining non-logging edits | 1,711 (10) | 3,687 (9) | 1,600 (10) | 340 (9) | 386 (8) | 211 (3) | 7,935 (49) |

## 4 Experimental Design

### 4.1 Evaluation Metrics

Following prior work on automated logging [34, 45, 74], we organize our evaluation around two high-level questions: *where to log* and *what to log*. The first dimension evaluates whether a model inserts a logging statement at the correct program location. The second evaluates whether a correctly placed statement uses the appropriate logging framework, severity level, dynamic variables, and message text. This separation is important because content quality is not meaningful when the predicted logging point is already incorrect. We therefore compute PA on all evaluated samples and report the remaining metrics only on the subset of position-correct predictions.

*4.1.1 where-to-log.* In line with previous work [34, 45], we employ **Position Accuracy (PA)** to assess the performance of logging position prediction. We also argue that the block level might be overly coarse [34]. Hence, we calculate PA as 1 (indicating a successful prediction) if the distance between the predicted line number and the actual line number is less than or equal to one line and both predicted and actual line numbers must be within the same block. Otherwise, PA is calculated as 0 (indicating an unsuccessful prediction).

*4.1.2 what-to-log.* On the position-correct subset $\mathcal{I}_{\text{PA}}$, we evaluate four complementary aspects of generated logging statement content.

(1) Logging framework anchor. In a multilingual benchmark, a model can place a logging statement at the correct location but still invoke a lexical receiver or macro that is inconsistent with the project convention. To capture this issue, we introduce **Framework Accuracy (FA)**, operationalized as exact matching of a strict lexical framework anchor:

$$\text{FA} = \frac{1}{M} \sum_{i \in \mathcal{I}_{\text{PA}}} \mathbf{1}(\hat{f}_i = f_i)$$

where $\hat{f}_i$ and $f_i$ denote the strict lexical anchors extracted from the predicted and gold logging statements, respectively. If no valid anchor can be extracted from a prediction, the sample is counted as incorrect. FA therefore measures lexical receiver/macro consistency rather than attempting to resolve the underlying library provider or semantic equivalence between APIs.

(2) Logging level. We evaluate log-level prediction using **Level Accuracy (LA)** and **Average Ordinal Distance (AOD)**, following prior studies on automated logging and log-level recommendation [40, 42]. LA measures exact agreement between predicted and gold levels:

$$\mathrm{LA} = \frac{1}{M} \sum_{i \in \mathcal{I}_{\mathrm{PA}}} \mathbf{1}(\hat{\ell}_i = \ell_i)$$

where $\hat{\ell}_i$ and $\ell_i$ are the predicted and gold logging levels for sample $i$. Because logging levels are ordinal rather than independent labels, we further use AOD to measure how close the predicted level is to the gold level under the evaluator's normalized severity order:

$$\mathrm{AOD} = \frac{1}{M} \sum_{i \in \mathcal{I}_{\mathrm{PA}}} \left(1 - \frac{\mathrm{Dis}(\ell_i, \hat{\ell}_i)}{\mathrm{MaxDis}(\ell_i)}\right)$$

where $\mathrm{Dis}(\ell_i, \hat{\ell}_i)$ is the ordinal distance between the predicted and gold levels, and $\mathrm{MaxDis}(\ell_i)$ denotes the maximum possible distance from the gold level $\ell_i$ within the level set.

(3) Logging variables. We evaluate whether the model captures the intended runtime information using **Precisely Matched Rate (PMR)** and set-based **Precision/Recall/F1** [33]. PMR measures the ratio of samples for which the predicted variable set exactly matches the gold variable set:

$$\mathrm{PMR} = \frac{1}{M} \sum_{i \in \mathcal{I}_{\mathrm{PA}}} \mathbf{1}(\hat{V}_i = V_i)$$

where $\hat{V}_i$ and $V_i$ are the predicted and gold variable sets for sample $i$. To capture partial matches, we compute set-based precision and recall for each sample and its harmonic mean:

$$\mathrm{Precision}_i = \frac{|\hat{V}_i \cap V_i|}{|\hat{V}_i|}, \qquad \mathrm{Recall}_i = \frac{|\hat{V}_i \cap V_i|}{|V_i|}, \qquad \mathrm{F1}_i = \frac{2 \cdot \mathrm{Precision}_i \cdot \mathrm{Recall}_i}{\mathrm{Precision}_i + \mathrm{Recall}_i}$$

We require exact matches at the variable-expression level after normalization. Thus, predictions that reuse the same base variable but access a different field or member function are treated as mismatches because they expose different runtime information. When both $\hat{V}_i$ and $V_i$ are empty, we treat the sample as a perfect variable match; when only one set is empty, the corresponding score is 0.

(4) Logging text. Finally, we evaluate generated logging messages using **BLEU-4** [49] and **ROUGE-L** [41], following prior work on automatic logging text generation [12, 45]. BLEU-4 measures local n-gram overlap, while ROUGE-L measures longest-common-subsequence similarity between the generated and gold messages. Higher scores indicate closer correspondence to developer-written logging text.

(5) Overall quality. In addition to the component-wise metrics above, we report a **Cascaded Composite Score (CCS)** for overall quality of generated logging statements. A simple arithmetic mean over PA, FA, and content metrics can be misleading in our setting, because FA, LA, message quality, and variable quality are only meaningful after the model first chooses the correct logging point. Moreover, even when the logging point is correct, selecting the wrong framework can still make the generated statement unusable in practice. We therefore design the composite score so that PA serves as the primary gate, FA as the secondary gate, and the remaining content quality as the tertiary component.

More generally, we define a residual content-quality term using one representative metric for each remaining aspect:

$$Q_\lambda = \lambda_{\text{LA}} \cdot \text{LA} + \lambda_{\text{R}} \cdot \text{ROUGE-L} + \lambda_{\text{F1}} \cdot \text{F1}, \qquad \lambda_{\text{LA}} + \lambda_{\text{R}} + \lambda_{\text{F1}} = 1$$

where LA represents level correctness, ROUGE-L represents message quality, and F1 represents variable recovery quality. We use these three representative metrics to avoid double-counting closely related measures such as AOD, BLEU-4, PMR, and variable-set precision/recall. The corresponding generalized CCS family is:

$$\text{CCS}_{w,\lambda} = \text{PA} \cdot \left(w_0 + w_{\text{FA}} \cdot \text{FA} + w_Q \cdot Q_\lambda\right). \qquad w_0 + w_{\text{FA}} + w_Q = 1$$

In the main results, unless otherwise stated, we set $w_0 = 0.5$, $w_{\text{FA}} = 0.25$, $w_Q = 0.25$, and $\lambda_{\text{LA}} = \lambda_{\text{R}} = \lambda_{\text{F1}} = 1/3$. This default setting preserves the priority structure of the task. The factor PA sets the overall upper bound of the score, ensuring that poor logging-point prediction cannot be compensated for by strong content quality on the smaller position-correct subset. Within the position-correct subset, FA receives the next-largest contribution because an incorrect framework often breaks compilability or practical usability. The remaining contribution is allocated evenly across level selection, message generation, and variable prediction through $Q_\lambda$. Because the exact weight allocation remains partly design-dependent, RQ1.4 later stress-tests the observed model-performance and model-ranking patterns under alternative hierarchy-preserving outer weights and alternative residual-quality allocations.

### 4.2 Studied LLMs

We select the studied models based on the nature of automated logging generation as a code-conditioned text generation task. A model for this task must not only understand program structure, control flow, and local execution context, but also generate concise and developer-like logging statements that fit different APIs and language ecosystems. We therefore choose a compact set of strong contemporary LLMs that covers both frontier closed-source models (**Claude-Sonnet-4.6** [2], **GPT-5.4** [48], and **Gemini-3.1-Pro** [61]), together with **DeepSeek-V3.2** [11], **GLM-5** [20], **Kimi-K2.5** [28], and **Qwen3-Coder-480b** [50] to broaden coverage across additional providers and model families, including code-specialized models. This selection gives us diversity in provider background, model family, and degree of code specialization while keeping the evaluation practical to run at our large-scale multilingual benchmark.

### 4.3 Implementation Details

For LLM inference, we query each studied model through its official API and normalize all returned outputs into a common structured format for downstream evaluation. We set the temperature to 0 for all models; we note that for some providers, temperature-0 inference does not guarantee fully deterministic outputs due to internal service variability [47]. For the prompting strategy, we adapt the validated best practice from previous studies [74]. Specifically, for each input sample from any of the three experimental branches, we employ BM25 [52] to retrieve the single most similar masked callable from the 6,426-instance repository-snapshot validation split. Thus, the revision-history and transformed revision-history data are used only for evaluation and do not provide exemplar candidates. The details of the prompt are shown in Listing 1.

For the transformed setting, the transformation pipeline first rematches each target logging statement in the original code using tree-sitter-based parsers, then applies a fixed set of target-preserving surface rewrites, including parenthesis insertion, condition-neutral rewriting, comment and docstring stripping, and redundant block normalization. Each transformed sample is re-parsed and re-validated; a rewrite is accepted only when the target logging statement remains unchanged and can still be rematched correctly in the transformed code.

Unless otherwise stated, we compute metrics at the sample level, aggregate them to repository-level means, and then report language-level macro averages so that large repositories do not dominate the final results. PA is averaged over all evaluated samples; the remaining metrics are averaged over the position-correct subset only. For cross-language summary rows, we further take the arithmetic mean over the six language-level values.

For experiments on the revision-history and transformed revision-history data, we run five repeated inference-and-evaluation rounds under the same prompt template, retrieval pool, evaluator, model set, and temperature setting. We report means across these rounds as point estimates and use the corresponding SD to characterize output variability. For RQ3 revision-history rank correlations and RQ4 transformed-minus-original deltas, we estimate finite-sample uncertainty with 10,000 repository-clustered bootstrap iterations: repositories are resampled with replacement, samples are resampled within each selected repository, and the same resampled cases are reused across models and repeated rounds to preserve paired comparisons.

Listing 1. Prompt template used.

```
System instruction:
You are a coding assistant that helps developers add appropriate logging statements to
    their code. The following function input misses a logging statement, please help me
    add a logging statement to the function to the appropriate place.

User prompt:
#example input:
<retrieved masked callable>
#example output:
{"revised_code": "<retrieved full callable with the gold logging statement restored>"}
#input code:
<target masked callable>
```

### 4.4 Research Questions

*RQ1. How does the performance of the evaluated models vary across programming-language ecosystem conditions?* Prior studies on automated logging generation have largely been conducted in single-language settings, making it unclear whether conclusions about model effectiveness generalize beyond the language ecosystem used for evaluation. RQ1 therefore treats each programming-language ecosystem as an evaluation condition and characterizes how the same models perform across the corresponding benchmark subsets. We decompose this comparison into four sub-questions:

- **RQ1.1:** What performance differences do the evaluated models exhibit across language-specific benchmark subsets?
- **RQ1.2:** Which logging-statement components show the largest within-model variation across language-specific benchmark subsets?
- **RQ1.3:** How stable are model rankings and model-selection conclusions across language-specific benchmark subsets?
- **RQ1.4:** Are the observed model-performance and model-ranking patterns robust to alternative CCS weightings?

*RQ2. How does the structural position of a logging statement within a function affect generation difficulty?* RQ2 shifts the analysis from language as the primary grouping variable to the structural position of the target logging statement within the enclosing function. The motivation is that two logging statements may belong to the same language yet pose very different prediction challenges depending on whether they occur in a top-level path, a branch body, a loop, an exception-handling region, or a nested callback-like scope [17, 39]. Unlike the language-level analyses in RQ1, this question asks whether there is a stable structure-linked difficulty profile that cuts across languages.

*RQ3. Which multilingual findings from repository-snapshot data persist in maintenance-oriented revision-history data?* Repository-snapshot data is well-suited for controlled, large-scale multilingual comparison. The strength of the resulting evidence, however, ultimately rests on whether the observed patterns persist when the prediction target is grounded in a developer's logging-statement introduction during code evolution. Revision-history data provides a distinct test of external validity: rather than predicting logging statements that already exist in frozen snapshots, models receive the parent-version callable and predict the logging statement introduced in the corresponding child revision. We therefore treat RQ3 not as a second multilingual leaderboard but as maintenance-oriented validation of conclusions established under the controlled repository-snapshot setting.

*RQ4. Do aggregate revision-history performance results remain robust under target-preserving surface transformations?* A concern with any benchmark drawn from public version histories is that strong model performance may partly reflect surface-form memorization rather than genuine logging reasoning. If a model encountered the same function bodies during pretraining, it could reproduce the original developer's logging decision without generalizing the underlying log-placement skill. This threat is especially salient in the revision-history setting, where code contexts are drawn from real open-source commits that may appear verbatim in pretraining corpora. RQ4 directly addresses this concern by asking whether aggregate revision-history performance changes broadly under a controlled surface perturbation that preserves the target logging decision while altering the surrounding code.

## 5 Experiment Results

### 5.1 How does the performance of the evaluated models vary across programming-language ecosystem conditions?

#### 5.1.1 *RQ1.1. What performance differences do the evaluated models exhibit across language-specific benchmark subsets?*

*Results.* Table 4 presents the main multilingual results, and the language-level "Average" rows summarize model outcomes under each language-specific evaluation condition. Averaged across the seven evaluated models, the Java subset yields the highest CCS (0.563), followed by Go (0.539), C# (0.495), Python (0.454), C++ (0.416), and JavaScript (0.383). The component metrics show that this subset ordering does not hold uniformly across generation decisions. The models achieve their highest average ROUGE-L on the Go subset (0.767) and their highest variable-recovery F1 on the C# subset (0.805). On the JavaScript subset, they achieve high average framework-anchor and level accuracy (FA 0.976 and LA 0.872) but lower position accuracy (PA 0.412). The lowest average FA occurs on the C++ subset (0.836), while the other subset means range from 0.889 for C# to 0.976 for JavaScript.

The $N_{\mathrm{PA}}$ column in Table 4 makes the size of each position-correct subset explicit. Across models, the JavaScript subset contains 96–177 position-correct predictions contributing to the conditional content metrics, compared with 873–1,360 for the Java subset. These counts describe the raw contributing sample sets rather than pooled denominators for the displayed repo-macro metrics. Thus, the high FA and LA values on JavaScript characterize only predictions that first localized the logging statement correctly, whereas the lower PA and CCS summarize the models' end-to-end outcomes over the full subset.

Table 4. Repo-macro core-benchmark results for all studied models across six language-specific subsets.

| Language | Model | Position | | Framework | Level | | Message | | Variables | | | | Overall |
|---|---|---|---|---|---|---|---|---|---|---|---|---|---|
| | | PA | $N_{\text{PA}}$ | FA | LA | AOD | BLEU-4 | ROUGE-L | PMR | Prec. | Rec. | F1 | CCS |
| Java | Claude-Sonnet-4.6 | 0.649 | 1068 | 0.940 | 0.783 | 0.914 | 0.481 | 0.642 | 0.561 | 0.739 | 0.727 | 0.718 | 0.593 |
| | DeepSeek-V3.2 | 0.556 | 939 | 0.905 | 0.680 | 0.858 | 0.409 | 0.587 | 0.501 | 0.699 | 0.669 | 0.666 | 0.494 |
| | Gemini-3.1-Pro | **0.823** | 1360 | **0.955** | **0.900** | **0.959** | **0.685** | **0.798** | **0.729** | **0.858** | **0.829** | **0.833** | **0.782** |
| | GPT-5.4 | 0.573 | 969 | 0.911 | 0.734 | 0.893 | 0.435 | 0.620 | 0.530 | 0.703 | 0.710 | 0.690 | 0.514 |
| | GLM-5 | 0.602 | 999 | 0.906 | 0.712 | 0.886 | 0.437 | 0.615 | 0.537 | 0.721 | 0.693 | 0.689 | 0.538 |
| | Kimi-K2.5 | 0.612 | 1056 | 0.919 | 0.751 | 0.898 | 0.454 | 0.628 | 0.540 | 0.714 | 0.715 | 0.698 | 0.552 |
| | Qwen3-Coder-480b | 0.529 | 873 | 0.920 | 0.669 | 0.859 | 0.408 | 0.581 | 0.503 | 0.691 | 0.666 | 0.662 | 0.470 |
| | *Average* | 0.620 | – | 0.922 | 0.747 | 0.895 | 0.473 | 0.639 | 0.557 | 0.732 | 0.716 | 0.708 | 0.563 |
| Python | Claude-Sonnet-4.6 | 0.554 | 871 | **0.950** | 0.770 | 0.902 | 0.452 | 0.637 | 0.560 | 0.680 | 0.675 | 0.668 | 0.505 |
| | DeepSeek-V3.2 | 0.427 | 703 | 0.861 | 0.706 | 0.870 | 0.384 | 0.583 | 0.506 | 0.649 | 0.659 | 0.643 | 0.374 |
| | Gemini-3.1-Pro | **0.715** | 1087 | 0.945 | **0.875** | **0.952** | **0.563** | **0.737** | **0.661** | **0.788** | **0.763** | **0.765** | **0.668** |
| | GPT-5.4 | 0.482 | 765 | 0.946 | 0.758 | 0.897 | 0.392 | 0.586 | 0.477 | 0.646 | 0.661 | 0.639 | 0.435 |
| | GLM-5 | 0.469 | 812 | 0.913 | 0.747 | 0.902 | 0.387 | 0.591 | 0.471 | 0.622 | 0.617 | 0.607 | 0.418 |
| | Kimi-K2.5 | 0.470 | 741 | 0.937 | 0.798 | 0.905 | 0.434 | 0.619 | 0.541 | 0.664 | 0.672 | 0.656 | 0.427 |
| | Qwen3-Coder-480b | 0.396 | 685 | 0.910 | 0.772 | 0.904 | 0.379 | 0.576 | 0.472 | 0.639 | 0.625 | 0.616 | 0.353 |
| | *Average* | 0.502 | – | 0.923 | 0.775 | 0.904 | 0.427 | 0.618 | 0.527 | 0.670 | 0.668 | 0.656 | 0.454 |
| Go | Claude-Sonnet-4.6 | 0.646 | 875 | 0.953 | 0.846 | 0.930 | 0.574 | 0.780 | 0.666 | 0.799 | 0.791 | 0.785 | 0.606 |
| | DeepSeek-V3.2 | 0.503 | 696 | 0.947 | 0.751 | 0.900 | 0.512 | 0.725 | 0.627 | 0.777 | 0.772 | 0.763 | 0.464 |
| | Gemini-3.1-Pro | **0.746** | 1079 | **0.982** | **0.900** | **0.953** | **0.672** | **0.857** | **0.795** | **0.892** | **0.869** | **0.873** | **0.719** |
| | GPT-5.4 | 0.542 | 712 | 0.918 | 0.811 | 0.920 | 0.528 | 0.735 | 0.583 | 0.729 | 0.752 | 0.729 | 0.498 |
| | GLM-5 | 0.557 | 775 | 0.928 | 0.807 | 0.921 | 0.541 | 0.762 | 0.634 | 0.804 | 0.787 | 0.783 | 0.517 |
| | Kimi-K2.5 | 0.594 | 820 | 0.932 | 0.797 | 0.908 | 0.557 | 0.771 | 0.646 | 0.786 | 0.787 | 0.776 | 0.551 |
| | Qwen3-Coder-480b | 0.453 | 606 | 0.930 | 0.783 | 0.902 | 0.523 | 0.742 | 0.653 | 0.804 | 0.770 | 0.774 | 0.418 |
| | *Average* | 0.577 | – | 0.941 | 0.813 | 0.919 | 0.558 | 0.767 | 0.658 | 0.799 | 0.790 | 0.783 | 0.539 |
| C++ | Claude-Sonnet-4.6 | 0.497 | 370 | 0.811 | 0.676 | 0.867 | 0.419 | 0.577 | 0.660 | 0.725 | 0.726 | 0.720 | 0.431 |
| | DeepSeek-V3.2 | 0.438 | 319 | 0.820 | 0.677 | 0.869 | 0.376 | 0.541 | 0.559 | 0.627 | 0.641 | 0.627 | 0.376 |
| | Gemini-3.1-Pro | **0.616** | 457 | **0.903** | **0.833** | **0.932** | **0.565** | **0.678** | **0.711** | **0.772** | **0.761** | **0.761** | **0.564** |
| | GPT-5.4 | 0.451 | 329 | 0.835 | 0.618 | 0.820 | 0.402 | 0.570 | 0.550 | 0.641 | 0.665 | 0.643 | 0.388 |
| | GLM-5 | 0.477 | 352 | 0.834 | 0.641 | 0.847 | 0.410 | 0.594 | 0.596 | 0.678 | 0.674 | 0.665 | 0.413 |
| | Kimi-K2.5 | 0.455 | 318 | 0.827 | 0.674 | 0.857 | 0.401 | 0.575 | 0.591 | 0.693 | 0.692 | 0.678 | 0.394 |
| | Qwen3-Coder-480b | 0.399 | 291 | 0.820 | 0.621 | 0.830 | 0.398 | 0.576 | 0.622 | 0.702 | 0.698 | 0.693 | 0.344 |
| | *Average* | 0.476 | – | 0.836 | 0.677 | 0.860 | 0.425 | 0.587 | 0.613 | 0.691 | 0.694 | 0.684 | 0.416 |
| JavaScript | Claude-Sonnet-4.6 | 0.512 | 136 | 0.941 | 0.815 | 0.929 | 0.515 | 0.654 | 0.651 | 0.736 | 0.731 | 0.730 | 0.470 |
| | DeepSeek-V3.2 | 0.315 | 96 | 0.989 | 0.858 | 0.959 | 0.425 | 0.586 | 0.405 | 0.621 | 0.606 | 0.592 | 0.289 |
| | Gemini-3.1-Pro | **0.607** | 177 | 0.978 | **0.988** | **0.991** | **0.717** | **0.839** | **0.784** | **0.815** | **0.805** | **0.808** | **0.585** |
| | GPT-5.4 | 0.389 | 111 | 0.969 | 0.832 | 0.936 | 0.458 | 0.642 | 0.562 | 0.652 | 0.665 | 0.651 | 0.358 |
| | GLM-5 | 0.341 | 98 | 0.963 | 0.889 | 0.946 | 0.560 | 0.642 | 0.601 | 0.726 | 0.669 | 0.687 | 0.315 |
| | Kimi-K2.5 | 0.389 | 116 | 0.990 | 0.857 | 0.942 | 0.485 | 0.644 | 0.527 | 0.574 | 0.559 | 0.562 | 0.358 |
| | Qwen3-Coder-480b | 0.331 | 103 | **1.000** | 0.868 | 0.941 | 0.417 | 0.594 | 0.568 | 0.671 | 0.653 | 0.652 | 0.307 |
| | *Average* | 0.412 | – | 0.976 | 0.872 | 0.949 | 0.511 | 0.657 | 0.586 | 0.685 | 0.670 | 0.669 | 0.383 |
| C# | Claude-Sonnet-4.6 | 0.580 | 381 | 0.910 | 0.829 | 0.936 | 0.498 | 0.705 | 0.698 | 0.816 | 0.794 | 0.795 | 0.535 |
| | DeepSeek-V3.2 | 0.489 | 323 | 0.878 | 0.760 | 0.888 | 0.498 | 0.682 | 0.675 | 0.839 | 0.815 | 0.811 | 0.444 |
| | Gemini-3.1-Pro | **0.709** | 433 | **0.956** | **0.907** | **0.969** | **0.643** | **0.798** | **0.805** | **0.900** | **0.865** | **0.874** | **0.676** |
| | GPT-5.4 | 0.508 | 331 | 0.831 | 0.838 | 0.927 | 0.445 | 0.661 | 0.670 | 0.829 | 0.818 | 0.808 | 0.457 |
| | GLM-5 | 0.529 | 346 | 0.862 | 0.802 | 0.918 | 0.488 | 0.660 | 0.668 | 0.807 | 0.776 | 0.780 | 0.478 |
| | Kimi-K2.5 | 0.505 | 330 | 0.876 | 0.823 | 0.915 | 0.503 | 0.669 | 0.686 | 0.808 | 0.792 | 0.788 | 0.459 |
| | Qwen3-Coder-480b | 0.458 | 311 | 0.911 | 0.697 | 0.889 | 0.504 | 0.659 | 0.676 | 0.821 | 0.771 | 0.783 | 0.415 |
| | *Average* | 0.540 | – | 0.889 | 0.808 | 0.920 | 0.511 | 0.690 | 0.697 | 0.831 | 0.804 | 0.805 | 0.495 |

Bold indicates the best score within each language block. Underlined bold indicates the best score in that metric across all language-model pairs, excluding the Average rows. $N_{\text{PA}}$ reports the raw number of position-correct predictions contributing to FA, LA, message, and variable evaluation in each language-model row. The displayed metric values remain repository-level macro averages and are therefore not pooled ratios over $N_{\text{PA}}$. Average rows summarize metric values across models and do not correspond to a single pooled $N_{\text{PA}}$.

This component pattern localizes much of the Java–JavaScript subset gap to target-site recoverability rather than logging-statement formatting alone. Because CCS is multiplicatively gated by PA, the higher PA obtained on the Java subset raises the models' end-to-end ceiling. Conversely, once the models correctly localize a sample in the JavaScript subset, their high conditional FA and LA indicate that they often select a matching framework anchor and severity level. The lower end-to-end outcomes on the JavaScript subset are therefore concentrated in logging-site localization within this benchmark. This interpretation diagnoses where model outcomes diverge but does not identify why: the qualitative cases and PA-failure decomposition in Section 5.5 provide case-level evidence about localization and realization errors, whereas model-training exposure and other sources of cross-condition variation remain unobserved.

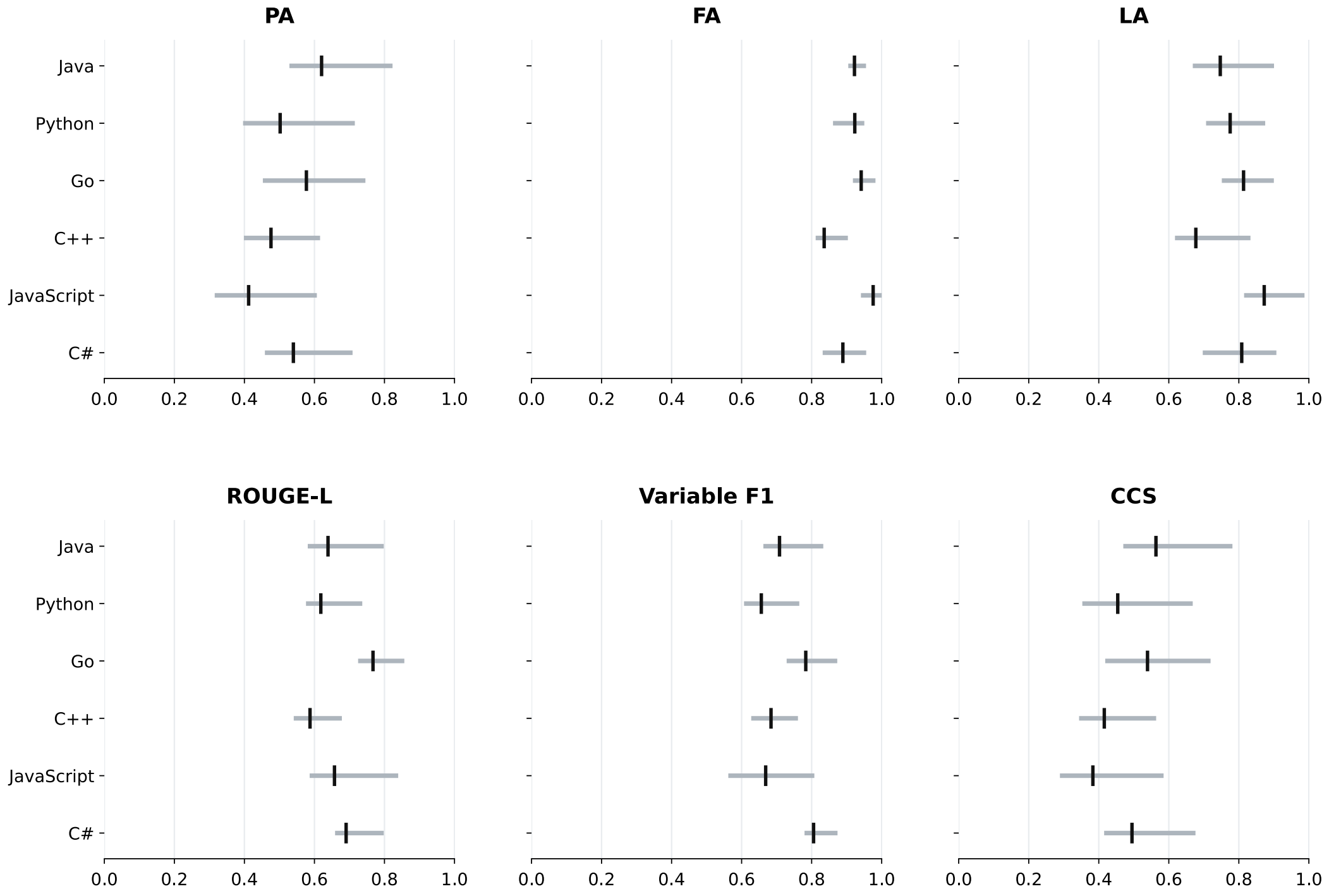

Fig. 3. Model-performance profiles across language-specific subsets of the core benchmark.

The same variation appears within individual models, showing that it recurs across evaluated systems rather than being driven by a single outlier model. For example, Gemini-3.1-Pro's PA changes from 0.823 on the Java subset to 0.607 on JavaScript, while its FA ranges from 0.982 on Go to 0.903 on C++. Similar within-model shifts appear in CCS: GPT-5.4 changes from 0.514 on Java to 0.388 on C++, and GLM-5 changes from 0.538 on Java to 0.315 on JavaScript. These comparisons establish variation in conditional model performance, but they do not isolate its causal source.

Figure 3 summarizes the same results as min–mean–max ranges across models, allowing us to assess whether the table-level pattern is stable or driven by a small number of outlier systems. The Java subset has the highest CCS band (0.470–0.782), and the models obtain a similar overall-quality band on Go (0.418–0.719) together with their highest ROUGE-L range (0.725–0.857). On C++, CCS ranges from 0.344 to 0.564 and FA from 0.811 to 0.903. The JavaScript subset has the lowest mean CCS despite a wider band (0.289–0.585): its FA and LA ranges are high at 0.941–1.000 and 0.815–0.988, but its PA range is only 0.315–0.607. These profiles show that model outcomes vary across language-specific subsets through different combinations of component-level strengths and weaknesses.

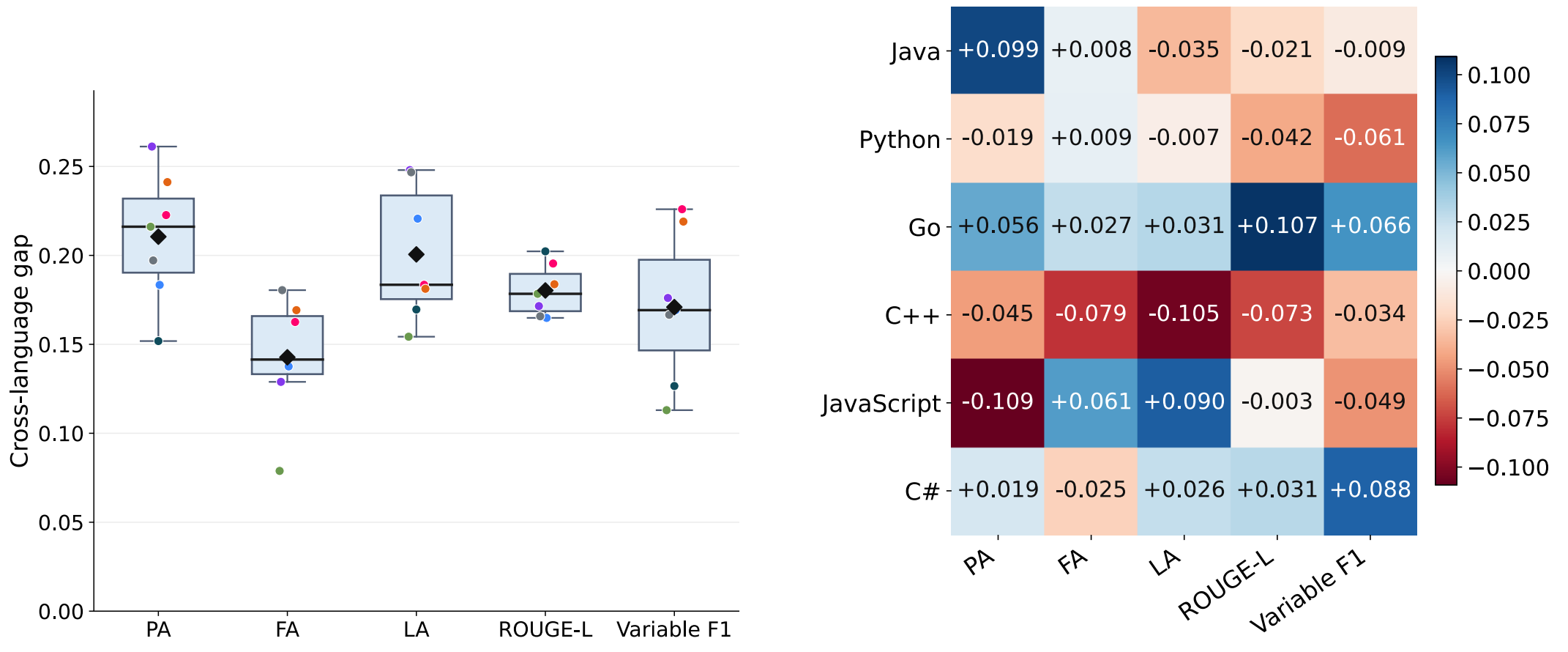

(a) Cross-language gap by representative metric.

(b) Language deviation from the multilingual mean.

Fig. 4. Cross-language component variation on five representative metrics.

**Finding 1**

The evaluated models exhibit clear performance variation across language-specific benchmark subsets under a shared protocol. Their highest mean CCS occurs on the Java subset, their highest message and variable-recovery scores occur on Go and C#, respectively, and their lowest mean CCS occurs on JavaScript, where high conditional framework-anchor and level accuracy do not offset lower position accuracy.

### *5.1.2 RQ1.2. Which logging-statement components show the largest within-model variation across language-specific benchmark subsets?*

*Results.* To answer this sub-question, Figure 4a shifts the analysis from overall scores to component families by computing, for each model and representative metric, the gap between its highest- and lowest-scoring language-specific subsets. The boxplots summarize variation across the seven evaluated models, and the overlaid points show the per-model gaps. Position prediction has the largest average within-model gap (PA 0.211), followed by level selection (LA 0.201), message quality (ROUGE-L 0.180), and variable recovery (F1 0.171). FA has the smallest average gap among these five representative metrics (0.143), with per-model gaps ranging from 0.079 to 0.180. Variation in model outcomes across the evaluation conditions is therefore distributed across the generation pipeline rather than being dominated by framework-anchor matching.

Having established the magnitude of within-model variation, Figure 4b shows where these differences occur by centering the cross-model mean for each language-specific subset and representative metric on the corresponding multilingual mean. Two contrasting profiles are especially informative. The JavaScript subset combines above-average conditional FA and LA with below-average PA and F1, separating relatively strong framework-anchor and level selection after correct placement from weaker logging-site localization and variable recovery. By contrast, outcomes on the C++ subset fall below the multilingual mean across multiple conditional content metrics, including FA, LA, ROUGE-L, and F1, showing that its lower realization outcomes extend beyond a single component after correct localization. These

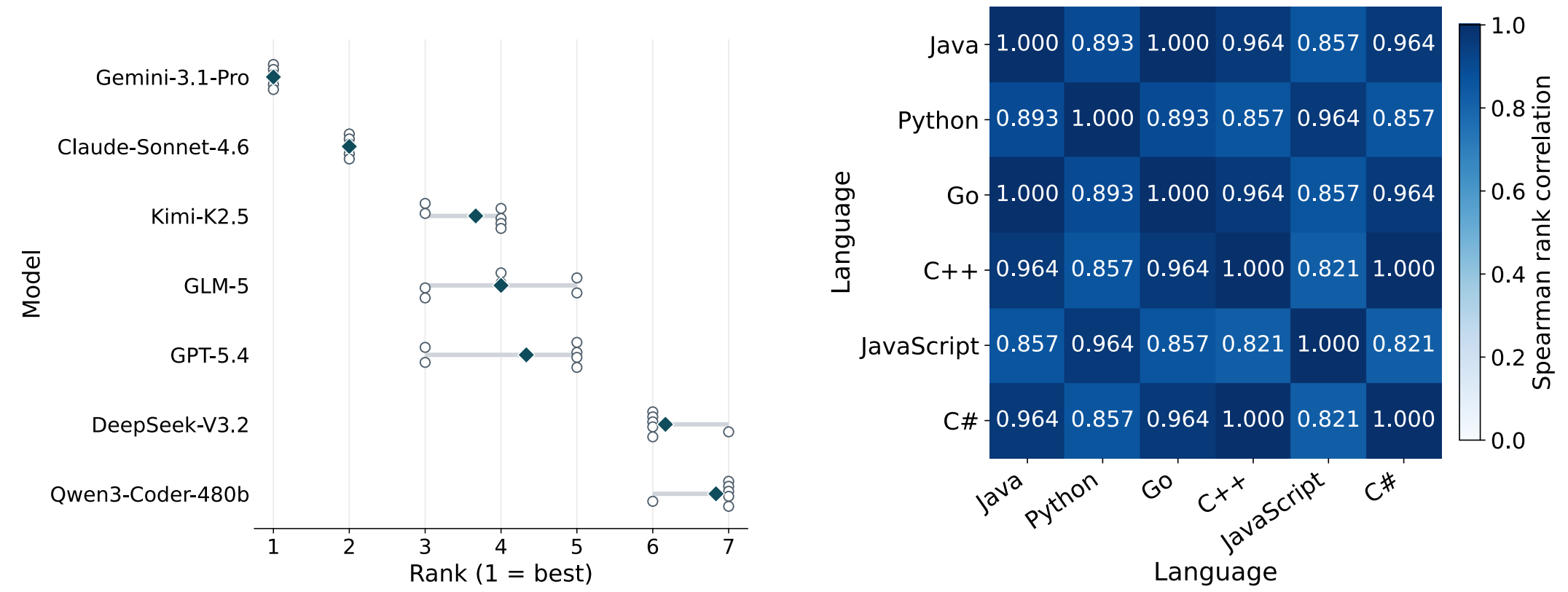

(a) Per-language model ranks.

(b) Pairwise rank correlation across languages.

Fig. 5. Cross-language model-ranking stability under CCS.

profiles show that cross-language outcome variation neither follows a single subset ordering across components nor concentrates in one generation decision.

The component-level view separates two forms of model-performance variation across evaluation conditions. First, PA dispersion reflects differences in how often models select the developer-authored logging site. This component is especially consequential because all downstream content metrics are conditioned on correct placement and CCS multiplies by PA. Second, the conditional content metrics characterize how well each model realizes the target logging decision after placement is correct. Together, these two views separate site localization from ecosystem- and project-conditioned realization, but they do not attribute either pattern to language alone.

**Finding 2**

Among the representative metrics, position prediction shows the largest average within-model dispersion across language-specific subsets, whereas strict framework-anchor matching shows the smallest. Model outcomes also exhibit distinct conditional component profiles: the JavaScript subset combines high framework-anchor and level accuracy with lower position accuracy, while outcomes on C++ are moderately below the multilingual mean across several content components.

#### 5.1.3 *RQ1.3. How stable are model rankings and model-selection conclusions across language-specific benchmark subsets?*

*Results.* Figure 5a ranks the seven models separately for each language and plots, for every model, its rank span together with the six language-specific rank positions. The language-specific leaderboards are highly but not perfectly aligned: the top two positions remain unchanged, while localized reordering occurs below them. Gemini-3.1-Pro and Claude-Sonnet-4.6 have zero-width spans and remain fixed at ranks 1 and 2 across all six languages, so multilingual evaluation does not alter the identity of the top two models. Below them, however, the ordering changes. Kimi-K2.5 is rank 3 in Java and Go but rank 4 in Python, C++, JavaScript, and C#; GLM-5 rises to rank 3 in C++ and C# but drops to rank 5 in Python and JavaScript; and GPT-5.4 reaches rank 3 in Python and JavaScript but remains at rank 5 in Java, Go, and C++. Even the bottom tier is not fixed: DeepSeek-V3.2 and Qwen3-Coder-480b swap ranks 6 and 7 in JavaScript. Thus, within this benchmark, a single-language leaderboard identifies the same top two models but can still change the recommended choice among non-dominant alternatives.

The middle-tier reordering is localized rather than a broad reversal of the leaderboard. Among Kimi-K2.5, GLM-5, and GPT-5.4, the model with the highest middle-tier CCS generally also has the highest PA: Kimi-K2.5 leads the three models on PA in Java and Go, GPT-5.4 does so in Python, and GLM-5 does so in C++ and C#. Because CCS multiplicatively gates the conditional content metrics by PA, variation in logging-site localization outcomes across language-specific subsets can change the ordering even when another model performs better on individual content components. JavaScript represents a particularly close case: GPT-5.4 and Kimi-K2.5 have the same rounded PA (0.389) and CCS (0.358), so their adjacent ordering should be interpreted as a near-tie rather than a substantial difference in capability.

The PA-failure decomposition in Section 5.5 provides complementary behavioral context. Multiple-logging-statement generation accounts for 54.0% of Kimi-K2.5's PA failures, whereas likely genuine misplacement accounts for 60.3% and 62.2% of the PA failures of GPT-5.4 and GLM-5, respectively. Thus, the middle-tier models do not fail position prediction in the same way: Kimi-K2.5 more often makes the target prediction non-unique, while GPT-5.4 and GLM-5 more often select a different code context. Together with their language-specific PA differences, these distinct failure profiles accompany the localized reordering among the middle-tier models across the six languages.

Figure 5b complements this analysis by reporting pairwise Spearman rank correlations between language-specific leaderboards. The mean pairwise Spearman correlation is 0.912, indicating high overall ranking agreement, although transferability is not uniform across language-specific subsets. Within this benchmark, the leaderboard computed on the JavaScript subset is the least reliable single-language surrogate for the others. Several pairs of subset-specific leaderboards are nearly or fully identical: Java and Go have a perfect correlation of 1.000, and C++ and C# also reach 1.000. Most remaining pairs are also highly correlated, such as 0.964 for Java versus C++, Java versus C#, and Python versus JavaScript, suggesting that the overall ranking trend is usually preserved. The lowest correlations, however, all involve the JavaScript-specific leaderboard, whose agreement drops to 0.857 with Java and Go and to 0.821 with both C++ and C#. This pattern is consistent with the rank-span analysis: evaluation on the JavaScript subset preserves the same top two models but is most likely to reshuffle the remaining models. The practical implication is therefore nuanced: single-language evaluation captures the leading models and the overall leaderboard trend in this benchmark, but it is not uniformly reliable for fine-grained cross-language model selection below the top two.

**Finding 3**

Model ranking is robust at the top but only partially transferable below it: Gemini-3.1-Pro and Claude-Sonnet-4.6 remain the strongest two models across all language-specific subsets, whereas variation in PA outcomes across subsets and distinct PA-failure profiles accompany localized reordering among the middle-tier models. Single-language leaderboards therefore remain useful for identifying the top two models in this benchmark, but are insufficient for fine-grained selection among closely matched alternatives.

#### *5.1.4 RQ1.4. Are the observed model-performance and model-ranking patterns robust to alternative CCS weightings?*

*Results.* Table 5 tests whether the model-performance and model-ranking patterns identified in RQ1 depend on the default CCS weighting. We examine 32 outer-weight variants satisfying $w_0 \geq w_{\text{FA}} \geq w_Q$ on a 0.05 grid, 16 residual-quality variants over LA, ROUGE-L, and F1, and 512 combined variants formed from their Cartesian product. Across the combined variants, Java retains the highest cross-model mean CCS, JavaScript retains the lowest, and the mean pairwise Spearman correlation across language-specific leaderboards remains between 0.905 and 0.924. Thus, the aggregate language-level and leaderboard-agreement conclusions are stable within the tested hierarchy-preserving weight space.

Table 5. Sensitivity of operationalized RQ1 conclusions under alternative CCS weightings on the core benchmark.

| Checked conclusion | Outer-weight sweep | Q-weight sweep | Combined sweep |
|---|---|---|---|
| Highest cross-model mean CCS remains on the Java subset | 32/32 (100.0%) | 16/16 (100.0%) | 512/512 (100.0%) |
| Lowest cross-model mean CCS remains on the JavaScript subset | 32/32 (100.0%) | 16/16 (100.0%) | 512/512 (100.0%) |
| Gemini-3.1-Pro remains rank 1 in every language-specific CCS ranking | 32/32 (100.0%) | 16/16 (100.0%) | 512/512 (100.0%) |
| Claude-Sonnet-4.6 remains rank 2 in every language-specific CCS ranking | 32/32 (100.0%) | 16/16 (100.0%) | 512/512 (100.0%) |
| Mean pairwise Spearman across CCS rankings remains at least 0.90 | 32/32 (100.0%) | 16/16 (100.0%) | 512/512 (100.0%) |
| At least one non-frontier model still changes rank across CCS rankings | 32/32 (100.0%) | 16/16 (100.0%) | 512/512 (100.0%) |

Table 6. Rank-level impact of the 512 combined CCS reweightings.

| Model | Default span | Sweep span | Changed subsets | Rank-level effect |
|---|---|---|---|---|
| Gemini-3.1-Pro | 1 | 1 | 0/6 | Fixed top-1 |
| Claude-Sonnet-4.6 | 2 | 2 | 0/6 | Fixed top-2 |
| Kimi-K2.5 | 3–4 | 3–5 | 2/6 | JavaScript/C#: adjacent swap with GPT-5.4 |
| GLM-5 | 3–5 | 3–5 | 0/6 | No weight-driven rank change |
| GPT-5.4 | 3–5 | 3–5 | 2/6 | JavaScript/C#: adjacent swap with Kimi-K2.5 |
| DeepSeek-V3.2 | 6–7 | 6–7 | 0/6 | Stable lower-tier placement |
| Qwen3-Coder-480b | 6–7 | 6–7 | 0/6 | Stable lower-tier placement |

Table 6 isolates the rank-level effect. Gemini-3.1-Pro and Claude-Sonnet-4.6 remain fixed at ranks 1 and 2, respectively. The only weight-induced changes are adjacent middle-tier swaps between Kimi-K2.5 and GPT-5.4 on the JavaScript and C# subsets. CCS therefore remains a robust summary of the broader ordering while the exact order of closely matched middle-tier models remains weight-sensitive.

**Finding 4**

The observed model-performance and model-ranking patterns are robust within the tested hierarchy-preserving CCS reweightings. Across all 512 variants, the evaluated models obtain the highest cross-model mean CCS on the Java subset and the lowest on JavaScript, while the top-two models remain unchanged across all language-specific rankings and leaderboard correlations stay high. Fine-grained rank changes are confined to adjacent middle-tier swaps between Kimi-K2.5 and GPT-5.4 on the JavaScript and C# subsets.

### 5.2 RQ2. How does the structural position of a logging statement within a function affect generation difficulty?

*Approach.* To address this question, we represent structural position using the structural context specified by the AST. Using the same syntax-aware detector as in dataset construction, we match each gold logging statement to its AST node within the enclosing callable. We then traverse the ancestor chain and assign the site to `nested_callable`, `exception`, `branch`, `loop`, or `top_level`, with the more specific `nested_callable` and `exception` labels taking precedence when multiple ancestor types are present. The non-top-level buckets should therefore be interpreted as enclosing structural contexts rather than literal coordinates in the function body. Moreover, `exception` denotes language-specific exception- or error-handling regions rather than a perfectly uniform construct across languages, and `nested_callable` serves as an operational bucket for callback, lambda, or local-function-like subscopes rather than a canonical language-independent category. For `top_level` logging statements, we further recover a coarse within-body position by locating the matched statement among the callable's direct body statements and partitioning that sequence into entry, middle, and exit thirds. These labels indicate source-order thirds of the top-level body, not formal control-flow entry or exit semantics. When exact body-child recovery is unavailable, we fall back to the statement's relative line ratio within the callable. For each bucket, we first compute language-specific repo-macro means for each model, then average across models and languages, and report the CCS gap as the maximum minus minimum CCS across languages.

Table 7. RQ2 results by structural bucket on the core benchmark. Metrics are multilingual means over language-specific repo-macro results.

| Structure | N | PA | FA | LA | ROUGE-L | Variable-F1 | CCS | CCS gap |
|---|---|---|---|---|---|---|---|---|
| Top-level (all) | 1228 | 0.429 | 0.805 | 0.764 | 0.561 | 0.579 | 0.383 | 0.273 |
| Entry | 485 | 0.366 | 0.621 | 0.621 | 0.432 | 0.491 | 0.323 | 0.352 |
| Middle | 352 | 0.426 | 0.694 | 0.665 | 0.486 | 0.486 | 0.384 | 0.306 |
| Exit | 391 | 0.410 | 0.679 | 0.593 | 0.510 | 0.508 | 0.371 | 0.145 |
| Branch | 3810 | 0.548 | 0.924 | 0.773 | 0.666 | 0.725 | 0.506 | 0.223 |
| Loop | 192 | 0.251 | 0.441 | 0.434 | 0.380 | 0.329 | 0.230 | 0.236 |
| Exception | 982 | 0.596 | 0.813 | 0.610 | 0.547 | 0.540 | 0.523 | 0.301 |
| Nested callable | 207 | 0.350 | 0.605 | 0.571 | 0.458 | 0.513 | 0.332 | 0.409 |
| *Average* | 6419 | 0.521 | 0.912 | 0.780 | 0.658 | 0.716 | 0.477 | 0.182 |

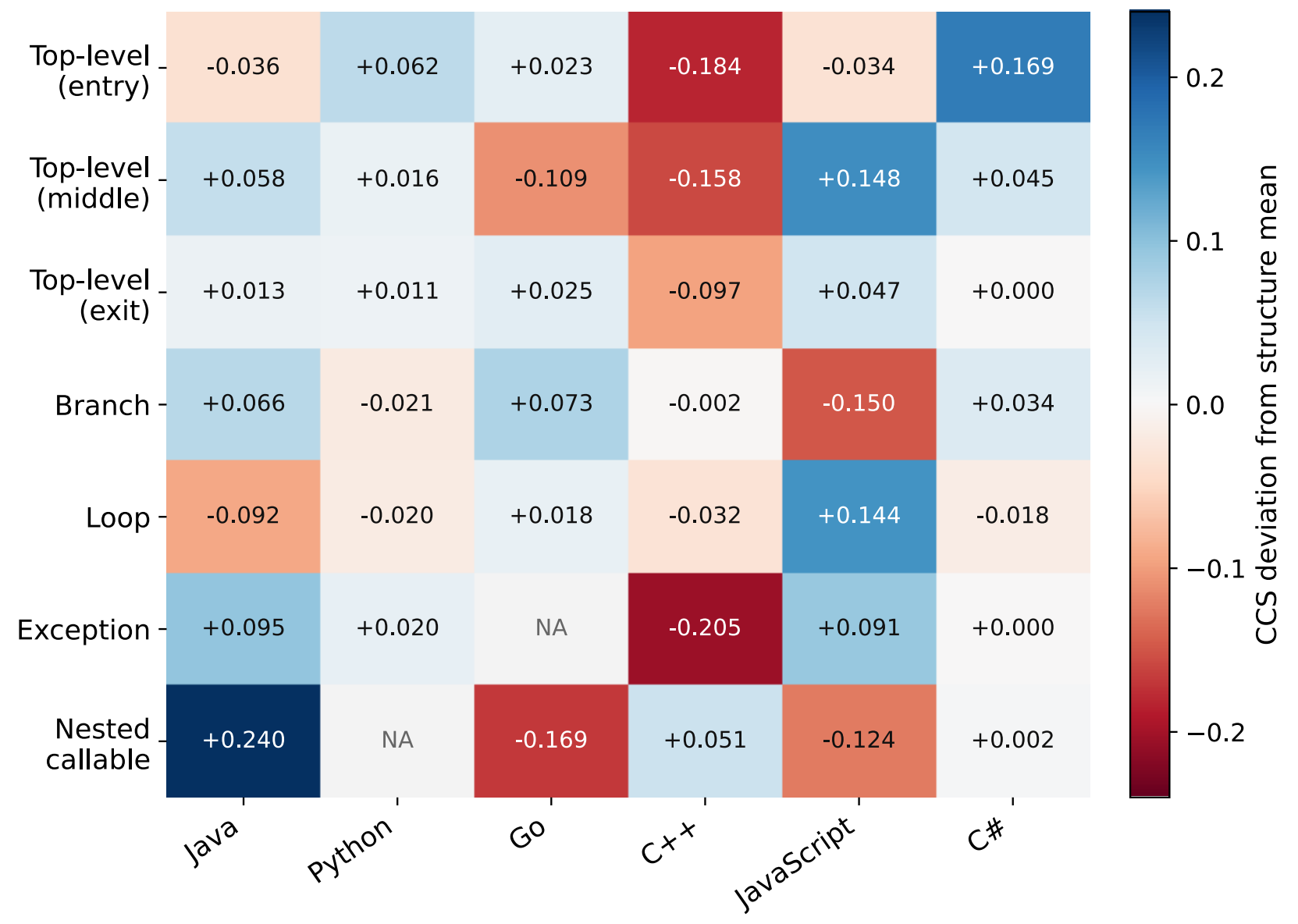

Fig. 6. Language deviation from each structure's mean CCS on the core benchmark. NA indicates that the corresponding slice has no test samples.

*Results.* Table 7 shows that the operationalized structural categories are associated with substantial differences in automated logging statement generation difficulty. The hardest bucket is `loop`, with a CCS of 0.230 and the lowest

PA (0.251), FA (0.441), and F1 (0.329) among the studied structures. Because FA and F1 are computed only on the position-correct subset, this pattern associates loop contexts with both lower position recovery and weaker content recovery once placement is correct, although it does not identify the cause of that difficulty. `nested_callable` is also challenging, with a CCS of 0.332. By contrast, `branch` and `exception` achieve the highest CCS among the major non-top-level buckets, at 0.506 and 0.523, respectively. Overall, the table supports a relative ordering in which loop and nested-callable contexts are harder than branch and exception contexts.

The aggregated `top_level` row reaches a CCS of 0.383, placing it near the middle of the overall difficulty range. Within this category, the entry slice is the hardest, with CCS 0.323 and PA 0.366, whereas the middle and exit slices reach CCS 0.384 and 0.371. Under our operationalization, earlier source-order top-level sites are therefore harder than later top-level sites on average. Because the entry/middle/exit split is only a coarse positional proxy, this result should not be over-interpreted as evidence about developers' cognitive stages or formal execution phases within a callable.

Figure 6 further shows that cross-language instability is structure-dependent rather than constant across buckets. Because the heatmap is centered on each bucket's mean CCS, it should be read as a relative deviation view rather than as a second ranking of absolute difficulty. Three patterns are most important. First, `nested_callable` exhibits the largest cross-language CCS gap (0.409): Java lies above the bucket mean, whereas Go and JavaScript lie below it. Second, a language's relative performance can reverse across structural contexts. JavaScript, for example, is above the `loop` mean (+0.144) but below the `branch` and `nested_callable` means (−0.150 and −0.124, respectively). Third, these reversals show that the aggregate difficulty ordering is not a fixed property of structural context independent of language-ecosystem conditions; rather, the relative difficulty associated with a structural context varies across those conditions.

A plausible explanation is that structural buckets differ in how explicitly they expose the logging purpose and the relevant runtime state. Branch and exception-handling regions often contain strong local cues, such as a failed condition, a state transition, a caught exception, or an error-return path. These cues narrow both the insertion point and the information that should be reported. Loop contexts are more ambiguous: a logging statement may describe per-iteration progress, an abnormal element, aggregate progress, or loop completion, and nearby placements can have different execution-frequency meanings. Nested-callable contexts add a scope-localization challenge because the model must distinguish the inner callback, lambda, or local-function-like scope from the enclosing callable and from adjacent nested handlers. The harder top-level entry slice is also consistent with a cue-sparsity interpretation: early source-order sites provide less preceding local context from which to infer whether the developer intended to report configuration, arguments, or initialization state. The qualitative structure cases in Section 5.5 illustrate these mechanisms at the case level, while the aggregate RQ2 results show that the same distinction appears systematically across the benchmark.

**Finding 5**

Under our AST-based operationalization of structural context, difficulty varies substantially across buckets: among the primary structural buckets, loop and nested-callable contexts are the hardest, while branch and exception-handling contexts are comparatively easy; within the top-level bucket, the earliest source-order slice is harder than the middle or exit slices. Cross-language instability is likewise context-dependent, with nested-callable contexts exhibiting the largest CCS gap. These differences are consistent with a context-localization explanation: branch and exception contexts expose stronger local cues, whereas loop and nested-callable contexts introduce target-state ambiguity and competing nearby scopes.

Table 8. Historical multilingual validation results for RQ3. Each cell reports mean±run-to-run SD; the colored subscript denotes the five-run revision-history mean minus the repository-snapshot value.

| Language | PA | FA | AOD | ROUGE-L | F1 | CCS |
|---|---|---|---|---|---|---|
| Java | $0.396 \pm 0.011_{-0.224}$ | $0.648 \pm 0.010_{-0.275}$ | $0.753 \pm 0.014_{-0.143}$ | $0.522 \pm 0.007_{-0.117}$ | $0.625 \pm 0.006_{-0.084}$ | $0.319 \pm 0.010_{-0.244}$ |
| Python | $0.510 \pm 0.009_{+0.007}$ | $0.830 \pm 0.023_{-0.093}$ | $0.862 \pm 0.009_{-0.042}$ | $0.516 \pm 0.005_{-0.103}$ | $0.578 \pm 0.009_{-0.078}$ | $0.436 \pm 0.010_{-0.018}$ |
| Go | $0.314 \pm 0.012_{-0.263}$ | $0.820 \pm 0.024_{-0.121}$ | $0.823 \pm 0.014_{-0.097}$ | $0.572 \pm 0.010_{-0.196}$ | $0.620 \pm 0.015_{-0.164}$ | $0.270 \pm 0.010_{-0.269}$ |
| C++ | $0.397 \pm 0.016_{-0.078}$ | $0.464 \pm 0.035_{-0.371}$ | $0.824 \pm 0.021_{-0.036}$ | $0.464 \pm 0.009_{-0.123}$ | $0.461 \pm 0.052_{-0.223}$ | $0.295 \pm 0.012_{-0.121}$ |
| JavaScript | $0.334 \pm 0.018_{-0.078}$ | $0.958 \pm 0.015_{-0.017}$ | $0.846 \pm 0.032_{-0.104}$ | $0.419 \pm 0.020_{-0.238}$ | $0.677 \pm 0.034_{+0.008}$ | $0.297 \pm 0.016_{-0.087}$ |
| C# | $0.307 \pm 0.023_{-0.233}$ | $0.540 \pm 0.044_{-0.349}$ | $0.773 \pm 0.028_{-0.147}$ | $0.530 \pm 0.010_{-0.160}$ | $0.721 \pm 0.033_{-0.085}$ | $0.240 \pm 0.019_{-0.255}$ |
| Mean | $0.376 \pm 0.015_{-0.145}$ | $0.710 \pm 0.025_{-0.205}$ | $0.813 \pm 0.020_{-0.095}$ | $0.504 \pm 0.010_{-0.156}$ | $0.614 \pm 0.025_{-0.104}$ | $0.309 \pm 0.013_{-0.166}$ |

### 5.3 RQ3. Which multilingual findings from repository-snapshot data persist in maintenance-oriented revision-history data?

*Approach.* To assess which aspects of the multilingual findings identified in RQ1 persist in a maintenance-oriented revision-history setting, we compare that setting directly with the shared six-language repository-snapshot data under the same evaluator and repo-level macro-aggregation protocol, using the seven LLMs evaluated in both settings. Following the repeated-evaluation protocol described in Section 4, Table 8 reports mean scores with SDs, and the colored delta denotes the revision-history mean minus the repository-snapshot value.

For model-selection transferability, we rank models within each language by mean CCS and compute pairwise Spearman correlations between the six language-specific leaderboards. We use the repository-clustered bootstrap interval from the repeated-evaluation protocol when interpreting the average revision-history rank correlation.

*Results.* Before examining which RQ1 findings persist, Table 8 establishes a setting-level performance shift. Compared with repository-snapshot data, the six-language mean is lower on all six displayed metrics, with the largest declines in FA, CCS, and ROUGE-L. One plausible explanation is the difference in target provenance: repository-snapshot evaluation masks a logging statement from a completed revision, whereas revision-history evaluation asks models to anticipate a logging statement subsequently introduced during maintenance. The latter therefore evaluates a maintenance-derived observability decision that was absent from the parent version.

Turning to the first RQ1 finding, cross-language score dispersion persists even though the subset ordering changes. Python becomes the highest-scoring subset under mean CCS, while Go and C# exhibit the largest CCS declines relative to repository-snapshot data. The spread among language-specific mean CCS values changes only modestly, from 0.180 to 0.196. Thus, the broad RQ1 conclusion that model outcomes vary materially across language-specific subsets still holds, although the relative difficulty of those subsets changes in the maintenance-oriented setting.

The component profile identified in RQ1 does not persist. In the revision-history setting, the average per-model FA gap increases from 0.143 to 0.518 and becomes the largest gap among the five representative component metrics used in RQ1.2. Table 8 further shows that the largest framework-anchor declines and the largest losses in other content metrics occur on different language-specific subsets. Thus, the maintenance-oriented setting exhibits substantially greater cross-language dispersion in framework-anchor matching rather than preserving the repository-snapshot component profile.

Figure 7a shows that another RQ1 conclusion weakens in the revision-history setting: model rankings show lower cross-language agreement. No model dominates all languages in the five-run-mean leaderboards. For example, Gemini-3.1-Pro leads on Python, Go, and C# but ranks last on C++, whereas GLM-5 leads on C++ but ranks last on Go.

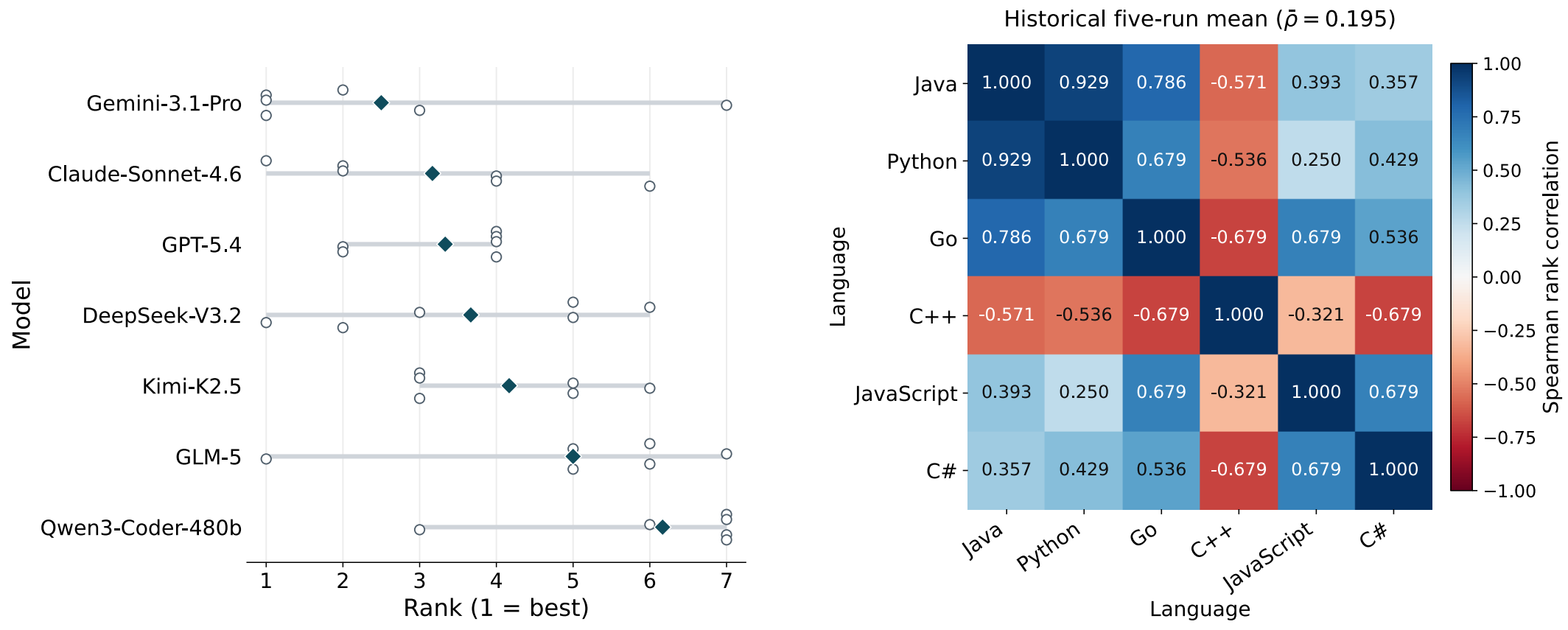

(a) Cross-language rank span by model.

(b) Pairwise rank correlation across languages.

Fig. 7. Cross-language model-ranking stability on revision-history data using mean CCS.

The revision-history setting therefore supports coarse-grained judgments about competitiveness, but not a single language-independent leaderboard for fine-grained model selection.

Figure 7b further quantifies this lower rank agreement. The average pairwise Spearman correlation across language-specific CCS leaderboards drops from 0.912 on repository-snapshot data to 0.195 on revision-history data. For the revision-history branch, the repository-clustered bootstrap 95% interval is $[-0.026, 0.400]$, and the run-to-run SD of the per-run mean correlations is 0.106. Taken together, these aggregate results indicate reduced fine-grained ranking transferability in the maintenance-oriented setting, while the wide uncertainty of individual language-pair estimates cautions against treating specific pairwise relationships as stable.

**Finding 6**

Revision-history data preserves cross-language score dispersion but not the repository-snapshot component profile or top-two ranking stability. The six-language mean is lower on every displayed metric, framework-anchor matching becomes the most dispersed component, and the mean pairwise Spearman correlation falls from 0.912 to 0.195.

### 5.4 RQ4. Do aggregate revision-history performance results remain robust under target-preserving surface transformations?

*Approach.* RQ4 checks whether the aggregate RQ3 performance results mainly come from models seeing the same code text during training. We run the same seven models on a transformed version of the revision-history benchmark. Following the repeated-evaluation protocol described in Section 4, we compute transformed-minus-original deltas within the same evaluation round, language, model, and metric. If exact surface-form memorization were the main reason for these performance results, the rewrites should lead to clear drops across languages and metrics.

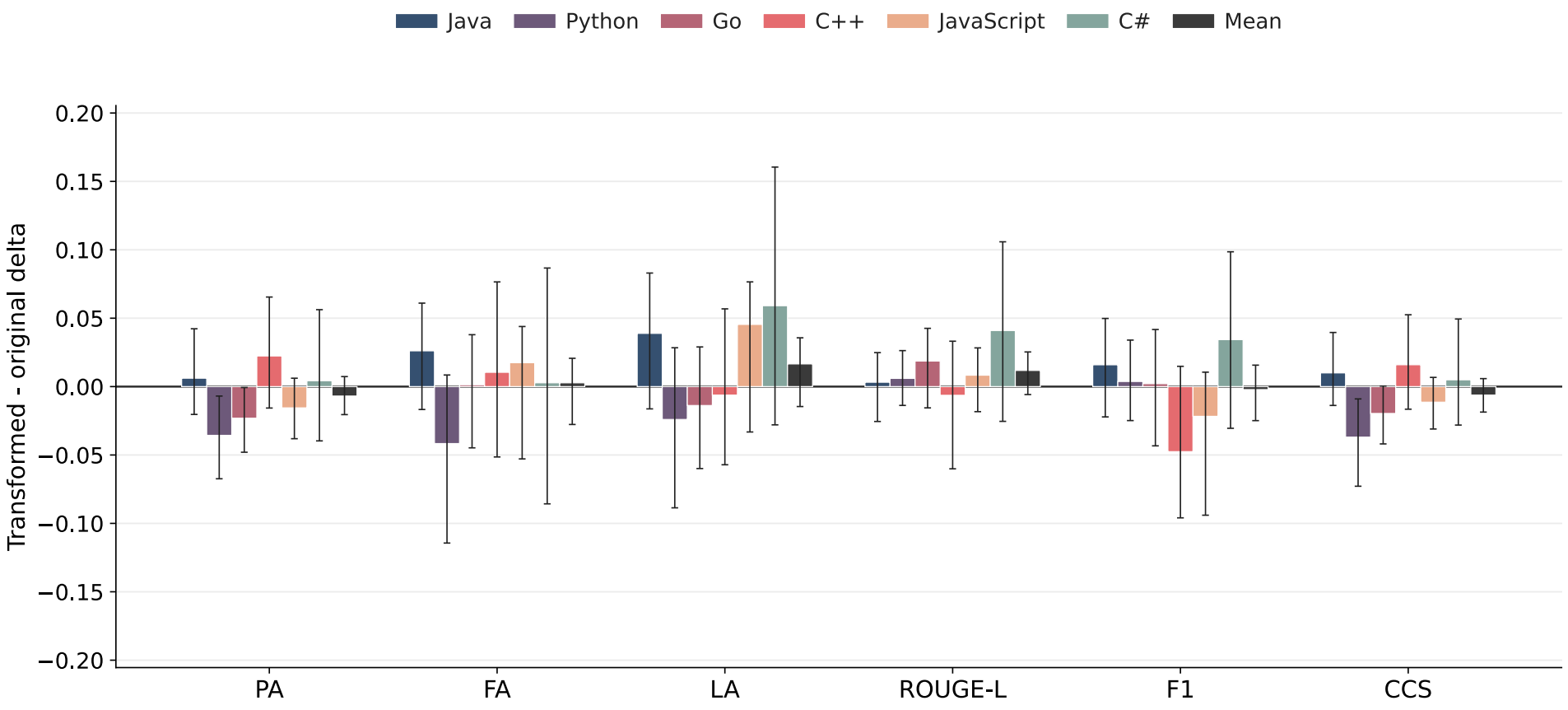

Fig. 8. RQ4 deltas between the transformed and original revision-history benchmarks. Bars show mean paired deltas; error bars show paired repository-clustered bootstrap 95% confidence intervals. *Mean* is the average over the six languages.

*Results.* Figure 8 shows that most changes after transformation are close to zero. The mean deltas are PA −0.007, FA +0.003, LA +0.017, ROUGE-L +0.012, F1 −0.002, and CCS −0.006. The SDs are also small: 0.021 for PA, 0.010 for FA, 0.017 for LA, 0.007 for ROUGE-L, 0.019 for F1, and 0.017 for CCS. The paired bootstrap 95% confidence intervals all include zero: PA $[-0.020, 0.007]$, FA $[-0.028, 0.021]$, LA $[-0.015, 0.036]$, ROUGE-L $[-0.006, 0.025]$, F1 $[-0.025, 0.016]$, and CCS $[-0.019, 0.006]$.

The language-level changes are small and go in both directions. Python has the largest CCS drop (−0.037), while C++ and Java have small CCS gains (+0.016 and +0.010). C++ has the largest F1 drop (−0.048), while C#, JavaScript, and Java improve on LA by +0.059, +0.045, and +0.039, respectively. This bidirectional pattern differs from the broad decline expected if exact code-text overlap were the primary driver of performance. Aggregate revision-history performance therefore does not exhibit a broad same-direction collapse after transformation. This analysis characterizes sensitivity to exact surface-form overlap; other forms of pretraining influence remain outside the transformation design.

**Finding 7**

The validated transformations do not lead to a clear drop in model performance across language-specific subsets: the mean FA and CCS changes are only +0.003 and −0.006, and both paired bootstrap intervals include zero. This pattern suggests that exact surface-form memorization is unlikely to be the main reason for the aggregate RQ3 performance results; more narrowly, those results do not exhibit a broad same-direction collapse under the transformations.

### 5.5 Qualitative Analysis of Failure Patterns

*Motivation.* The aggregate analyses in RQ1 and RQ2 identify two quantitative patterns: cross-language variation is distributed across generation components, with position prediction showing the largest average gap, while loop and nested-callable sites are the most challenging among the primary structural buckets. Aggregate metrics do not explain the local errors that produce these differences. We therefore examine representative cases from the shared

**C++ input context**

```
auto hosts = followers();
auto it = std::find_if(hosts.begin(), hosts.end(), [&candidate](const auto& h)
    {
  return h->address() == candidate;
});
if (it == hosts.end()) {
  VLOG(2) << idStr_ << "The candidate " << candidate // ground truth
          << " is not in my peers";
  return nebula::cpp2::ErrorCode::E_RAFT_INVALID_PEER;
}
```

**Representative prediction (GPT-5.4)**

```
LOG(WARNING) << "Candidate peer not found in current followers, "
             << "candidate: " << candidate;
```

Fig. 9. C++ framework confusion: `VLOG(2)` is reduced to generic `LOG(...)` forms.

repository-snapshot benchmark to characterize how models can recover the reporting intent of a site while still missing project- or context-specific realization constraints.

*Framework confusion across language ecosystems.* Models often recover the general reporting intent of a site but fail to preserve project-specific framework conventions. In `vesoft-inc/nebula`, the gold C++ statement uses `VLOG(2)`, whereas all seven models reduce it to generic forms such as `LOG(INFO)` or `LOG(WARNING)` (Figure 9). Likewise, in `dotnet/roslyn`, six models replace the static receiver `Logger` with the instance-level `_logger` (Figure 10). These failures are not primarily message-generation errors; rather, they reflect misalignment with the local logging framework. The predictions preserve much of the reporting intent, but they instantiate that intent with the wrong API family or receiver convention. Together, these cases illustrate how FA can deteriorate even when coarser semantic alignment remains relatively strong.

This pattern is not universal. In `alibaba/canal`'s Java method `startMonitor`, all seven models correctly recover `logger.error("scan remote adapter configs failed", e);`. This contrast is more consistent with the view that framework mismatch becomes more likely when the local codebase does not expose a single dominant convention that is easily recoverable from nearby context, rather than from the mere existence of framework choice itself.

**Finding 8**

Framework confusion is a salient local failure pattern even though FA does not show the largest aggregate within-model variation across language-specific subsets. The representative cases show that models can preserve the reporting intent of a logging statement while still selecting a project-inconsistent macro family or receiver convention.

*Structural ambiguity in loop and nested-callable contexts.* Loop and nested-callable sites exhibit a different failure mode. In the inspected cases, the dominant issue is context-localization difficulty rather than framework mismatch. Because RQ2 operationalizes these labels as AST-based enclosing structural contexts rather than formal control-flow

**C# input context**

```
public virtual bool TryGetLanguageForRequest(string methodName, object?
    serializedRequest,
        [NotNullWhen(true)] out string? language)
{
    // ground truth
    Logger.LogDebug("Using default language handler for {methodName}");
    language = LanguageServerConstants.DefaultLanguageName;
    return true;
}
```

**Representative prediction (GPT-5.4)**

```
_logger.LogDebug("Resolving language for request method {MethodName}",
                methodName);
```

Fig. 10. C# receiver mismatch: static Logger is replaced with instance-level _logger.

states, the discussion here should be read as a case-based interpretation of how models localize logging statements within those buckets. We examine three contrasting cases to clarify the error patterns underlying the structural pattern observed in RQ2.

In `apache/airflow`'s Python method `rewrite`, the gold logging statement appears inside a `while` loop and reports per-iteration progress (Figure 11). Models generate plausible progress-related messages, but they systematically misplace them, attaching the logging statement to the broader copy operation rather than to the loop-local update. The difficulty is therefore not message wording alone, but correct localization within the iterative subcontext.

The contrast with branch contexts is instructive. In `ray-project/ray`'s Python method `autoscale`, the gold statement reports an explicit upscaling transition (Figure 12). All seven models recover this logging statement exactly. The branch condition `new_num > old_num` provides an explicit state-change cue that tightly constrains both the insertion point and the message content. This case therefore offers a plausible explanation for why branch sites are easier in the quantitative results from RQ2.

Nested callbacks introduce a stronger form of ambiguity. In `kubernetes/kubernetes`'s Go method `NewController`, the gold statement belongs to the `AddFunc` callback (Figure 13). None of the seven models produce the correct statement in the correct callback. Some fall back to outer-scope initialization messages, whereas others copy semantically nearby patterns from adjacent handlers, such as "New pod" from the surrounding `UpdateFunc`. Here the prediction problem is not only to identify relevant event content, but also to align that content with the correct nested subscope among several adjacent callable blocks. These nearby callbacks provide distractor patterns that are locally plausible yet structurally incorrect, amplifying the localization challenge beyond what a single loop introduces.

Taken as explanatory cases rather than standalone proof, these examples suggest a pattern consistent with the quantitative results from RQ2: branch contexts provide explicit state-change cues and are recovered exactly in the inspected case; loop contexts make loop-local placement easier to miss; nested-callable contexts add competing nearby subscopes that increase the risk of misalignment.

**Python loop context**

```
while token is not None:
    token, bytes_rewritten, total_bytes = destination_bucket.blob(
        blob_name=destination_object
    ).rewrite(source=source_object, token=token)

    self.log.info("Total Bytes: %s | Bytes Written: %s",
                  total_bytes, bytes_rewritten) # ground truth
get_hook_lineage_collector().add_input_asset(
    context=self,
```

**Representative prediction (GPT-5.4)**

```
self.log.info("Rewriting object %s in bucket %s to object %s in bucket %s",
              source_object, source_bucket,
              destination_object, destination_bucket)
```

Fig. 11. Loop-local placement ambiguity: the iterative update is replaced by a broader operation-level message.

**Python branch context**

```
new_num = self._target_state.target_num_replicas
if new_num > old_num:
    logger.info(  # ground truth
        f"Upscaling {self._id} from {old_num} to {new_num} replicas. "
        f"{curr_stats_str}"
    )
    self._curr_status_info = self._curr_status_info.handle_transition(
```

**Representative prediction (GPT-5.4)**

```
logger.info(
    f"Upscaling {self._id} from {old_num} to {new_num} replicas. "
    f"{curr_stats_str}"
)
```

Fig. 12. Branch-local transition recovered exactly under an explicit state-change cue.

**Finding 9**

The qualitative cases indicate that differences across structural buckets are associated with different forms of context-localization error. Loop sites can blur whether a logging statement should describe the iterative update or the surrounding operation. Nested-callable sites can introduce neighboring handlers that provide competing yet structurally incorrect targets. These patterns are consistent with the RQ2 result that loop and nested-callable sites are harder than branch sites in the cross-language quantitative analysis.

**Go nested-callback context**

```
if _, err := claimInformer.Informer().AddEventHandlerWithOptions(cache.
    ResourceEventHandlerFuncs{
    AddFunc: func(obj interface{}) {
        logger.V(6).Info("New␣claim", "claimDump", obj) //ground truth
        ec.enqueueResourceClaim(logger, nil, obj)
    },
    UpdateFunc: func(old, updated interface{}) {
        logger.V(6).Info("Updated␣claim", "claimDump", updated)
        ec.enqueueResourceClaim(logger, old, updated)
    }
```

**Representative prediction (Gemini-3.1-Pro)**

```
logger.V(6).Info("New pod", "podDump", obj)
```

Fig. 13. Nested-callback distractor pattern: a neighboring handler provides a plausible but incorrect target.

Table 9. Rule-based decomposition of PA-failed cases on the repository-snapshot benchmark. Percentages are computed within each model's PA-failed cases.

| Model | PA failures | Output/extraction | Multiple logging statements | Near-target shift | Likely genuine misplacement |
|---|---|---|---|---|---|
| Gemini-3.1-Pro | 1,826 | 451 (24.7%) | 319 (17.5%) | 55 (3.0%) | 1,001 (54.8%) |
| Claude-Sonnet-4.6 | 2,718 | 279 (10.3%) | 719 (26.5%) | 39 (1.4%) | 1,681 (61.8%) |
| GPT-5.4 | 3,202 | 233 (7.3%) | 1,006 (31.4%) | 32 (1.0%) | 1,931 (60.3%) |
| GLM-5 | 3,037 | 364 (12.0%) | 754 (24.8%) | 30 (1.0%) | 1,889 (62.2%) |
| Kimi-K2.5 | 3,038 | 382 (12.6%) | 1,640 (54.0%) | 20 (0.7%) | 996 (32.8%) |
| DeepSeek-V3.2 | 3,343 | 500 (15.0%) | 960 (28.7%) | 28 (0.8%) | 1,855 (55.5%) |
| Qwen3-Coder-480b | 3,550 | 617 (17.4%) | 1,434 (40.4%) | 18 (0.5%) | 1,481 (41.7%) |
| All models | 20,714 | 2,826 (13.6%) | 6,832 (33.0%) | 222 (1.1%) | 10,834 (52.3%) |

*Rule-based decomposition of PA failures.* Because PA acts as a multiplicative gate in CCS, we further examine what kinds of errors are counted as PA failures. We decompose all negative cases into four mutually exclusive categories using a conservative rule-based taxonomy. The categories are applied in priority order. First, output/extraction failures denote cases where no comparable predicted logging statement is available, either because no changed logging statement was detected, or the extracted statement has no callable-relative line. Second, multiple logging statements denote cases where the model generates more than one changed logging statement in the target callable, making the intended prediction non-unique. Third, near-target line shifts denote cases with exactly one extracted logging statement whose line misses the PA tolerance but remains within five lines of the gold line. We require the predicted statement either to exactly match the normalized gold statement or to satisfy a conservative content-similarity rule: matching framework and level ($FA = 1$ and $LA = 1$), with ROUGE-L $\geq 0.8$ for the message text and variable $F1 \geq 0.8$. These cases capture local placement offsets rather than clear semantic relocation. Finally, likely genuine misplacement denotes cases with exactly one extracted logging statement that is placed on the wrong line and is not a near-target shift, suggesting that the model selected a different code context for the logging statement.

Table 9 shows that PA failures are not dominated by a single artifact. Across all models, likely genuine misplacement is the largest category, accounting for 10,834 of 20,714 PA failures (52.3%). Multiple logging statements form the second-largest category (6,832, 33.0%), followed by output/extraction failures (2,826, 13.6%) and near-target line shifts (222, 1.1%).

This distribution indicates that PA aggregates several distinct behaviors: true localization errors, multiple-statement generation, local line-offset errors, and output/extraction artifacts.

The same decomposition helps explain the Gemini-3.1-Pro result. Gemini-3.1-Pro has fewer PA failures than Claude-Sonnet-4.6 overall (1,826 vs. 2,718), with fewer likely genuine misplacements (1,001 vs. 1,681) and fewer multiple-logging-statement cases (319 vs. 719), although it has more output/extraction failures (451 vs. 279). Thus, Gemini-3.1-Pro's PA advantage is not merely a cleaner-output or extraction artifact. It is mainly associated with fewer localization and multiple-statement generation failures under the evaluator, while output/extraction behavior remains a non-negligible source of PA loss. This finding complements the CCS sensitivity analysis in RQ1: CCS can amplify PA differences through its multiplicative structure, but the observed Gemini lead is tied to several PA subcomponents rather than to formatting or extraction behavior alone.

**Finding 10**

PA failures are heterogeneous rather than a single formatting artifact. Across all PA-failed cases, likely genuine misplacement is the largest source of error (52.3%), followed by multiple logging statements (33.0%), output/extraction failures (13.6%), and near-target line shifts (1.1%). Gemini-3.1-Pro's PA advantage is mainly associated with fewer likely genuine misplacements and fewer multiple-logging-statement cases, indicating that CCS amplifies differences in localization behavior rather than merely differences in output formatting or extraction success.

## 6 Discussion

### 6.1 Lessons Learned and Practical Implications

We distill six lessons from RQ1–RQ4 and the qualitative failure analysis. Each lesson translates an empirical pattern into a bounded implication for future benchmark design and tool development.

*Model performance varies across language-ecosystem conditions.* RQ1 shows that variation in model outcomes across language-specific subsets involves both target-site recoverability and ecosystem-conditioned realization. Figure 4a reports the largest average within-model gap for position prediction (0.211), making logging-site localization the most variable evaluated component across these conditions. Figure 4b further shows distinct conditional content profiles after correct placement, while the representative cases in Section 5.5 illustrate how models can preserve a site's general reporting purpose yet select a project-inconsistent macro family or receiver convention. Such predictions may require developer correction even when their coarse reporting intent is plausible. Because the study does not separately control language syntax, project composition, logging conventions, or model-training exposure, these results establish condition-specific performance variation rather than intrinsic language difficulty. Single-language evaluations may therefore overestimate the transferability of absolute performance levels and component-specific weaknesses because they do not jointly evaluate localization and ecosystem-conditioned realization.

*Multilingual evaluation has claim-dependent value for model selection.* The repository-snapshot leaderboards exhibit high overall agreement: Figure 5 reports a mean pairwise Spearman correlation of 0.912 and the same top two models across all six language-specific subsets. A single-language leaderboard is therefore informative for identifying the leading models in this controlled benchmark. It does not, however, preserve the exact ordering of middle- and lower-tier models or predict their absolute performance under every language-ecosystem condition. In the revision-history setting, Figure 7 shows that the mean pairwise correlation decreases to 0.195 and that no model leads all six language-specific subsets. The repository-clustered bootstrap interval for the mean correlation, $[-0.026, 0.400]$, cautions against treating individual pairwise relationships as stable, but the setting-level results still show substantially weaker fine-grained ranking transferability. When a deployment decision depends on choosing among non-dominant alternatives or estimating performance for a specific language stack, candidate models should therefore be validated under the target ecosystem and maintenance condition.

*Grounding failures span multiple generation components.* The component breakdown in Figure 4 shows that cross-language variation spans logging-site localization, framework-anchor matching, severity prediction, message construction, and variable recovery. The PA-failure decomposition in Table 9 further shows that failed placement is not predominantly a local line-offset artifact: likely genuine misplacement is the largest category, accounting for 52.3% of all PA failures. The representative cases in Section 5.5 complement these aggregate results by illustrating project-inconsistent macro or receiver selection and context-localization errors. Together, the evidence indicates that automated logging challenges extend beyond surface message wording. Future systems should therefore account jointly for structural grounding, strict framework-anchor consistency, severity selection, message quality, and variable recovery, while evaluating whether improvements to one component introduce regressions in another.

*Deployment-oriented evaluation should check framework-anchor consistency separately from localization.* Framework-anchor matching is comparatively stable across language-specific subsets in the repository-snapshot benchmark, where Figure 4a reports the smallest average within-model gap among the representative metrics (0.143). In the revision-history setting, however, the average per-model FA gap increases to 0.518, even though FA is evaluated only after a prediction has correctly localized the logging site. This contrast shows that correct placement does not ensure agreement with the lexical receiver or macro in the developer-authored reference, particularly in the evaluated maintenance-oriented setting. Because FA measures strict lexical anchor consistency rather than underlying library identity or semantic equivalence between APIs, the supported practical implication is correspondingly specific: deployment-oriented evaluations should verify whether a generated receiver or macro follows the target repository's convention instead of treating correct localization as sufficient. We elaborate concrete ecosystem-aware generation and verification directions in the Future Work subsection.

*Structural difficulty is consistent with local-purpose and target-state ambiguity.* Table 7 shows that loop and nested-callable contexts are harder than branch and exception contexts under the study's AST-based structural operationalization. These aggregate associations establish a difficulty ordering but do not by themselves identify its cause. The representative cases in Section 5.5 provide a consistent explanation: the inspected loop case requires localization to an iterative update, the nested-callable case contains competing adjacent subscopes, and the branch case exposes an explicit state-change cue. The combined evidence therefore motivates structure- and scope-aware generation methods that recover the local reporting purpose and relevant runtime state. Local-purpose and target-state ambiguity should nevertheless be interpreted as an evidence-consistent explanation of the observed structural pattern rather than an established causal mechanism.

*Controlled benchmarks require realism checks for engineering claims.* The repository-snapshot benchmark exposes variation in model outcomes across language-specific subsets, including absolute score gaps and component-level dispersion. Table 8 shows that revision-history evaluation preserves cross-language score dispersion but yields a lower six-language mean on every displayed metric, while Figure 7 shows that the repository-snapshot top-two stability does not persist. The two settings therefore support different claims: controlled multilingual evaluation identifies model-performance patterns under repository-snapshot masking, whereas maintenance-oriented evaluation tests which of those findings transfer to logging-statement-introducing changes derived from revision histories. Strong repository-snapshot performance alone is insufficient evidence that the same absolute performance levels, component profiles, or model rankings transfer to maintenance-derived cases.

## 6.2 Threats to Validity

### 6.2.1 Construct validity.

*Reference-answer ambiguity.* Each benchmark instance uses a developer-authored logging statement as the reference, but that statement is not necessarily the only valid answer [33]. Multiple message wordings, variable selections, and nearby logging sites may all be reasonable for the same context. Our evaluator may therefore underestimate absolute performance when models produce plausible but non-matching alternatives. We mitigate this risk by reporting a multi-view metric suite and framing claims around comparative cross-language trends rather than single-metric assessments of logging quality.

### 6.2.2 Internal validity.

*Benchmark construction and filtering.* Benchmark construction relies on automated mining, parsing, callable alignment, and normalization, so extraction noise may remain despite syntax-aware analysis and strict cleaning. This risk may be uneven across languages because logging APIs, macro styles, parser behavior, and framework conventions differ. The revision-history branch focuses on local logging-statement-introducing edits with limited non-logging change, which favors cleaner maintenance cases and may underrepresent revisions in which logging changes accompany broader behavioral modifications. Syntax-aware detection, callable-level alignment, strict cleaning, and conservative commit filtering reduce extraction noise while defining the scope of the accepted pool. The resulting evidence applies to relatively clean, local logging-statement-introducing edits rather than the full range of logging-related maintenance.

*Attribution of variation across language-specific subsets.* Cross-language outcome differences may jointly reflect programming-language syntax, logging APIs and conventions, project and sample composition, retrieved-exemplar quality, and the quantity or quality of relevant language data in model pretraining or post-training. The shared prompt template, BM25 exemplar pool, inference configuration, evaluator, and repository-level macro aggregation hold the evaluation pipeline constant, but these potential sources of variation are not independently controlled. RQ1 therefore characterizes conditional model performance and the transferability of model-selection conclusions across language ecosystems rather than intrinsic language difficulty.

*Data leakage and pretraining overlap.* Because our benchmark instances are mined from public repositories and revision histories, some inputs and developer-authored targets may have appeared in the pretraining or fine-tuning corpora of the evaluated models. Prior code-generation studies have found benchmark overlap with open training corpora and higher performance on contaminated subsets, demonstrating that such overlap can inflate apparent generalization [46, 51]. RQ4 measures sensitivity to exact and near-exact surface-form overlap by evaluating the same 744 revision-history cases after conservative, target-preserving transformations: 625 cases (84.0%) are rewritten, whereas 119 remain identity fallbacks. Across five repeated paired runs, the transformed branch exhibits no broad same-direction decline, with a mean CCS change of $-0.006$ and a paired bootstrap 95% CI of $[-0.019, 0.006]$. This stability reduces the plausibility of verbatim recall as the primary explanation for the aggregate RQ3 results. The local rewrites preserve the gold target and may retain recognizable semantic patterns and repository conventions, while the identity fallbacks retain their original form. Other forms of pretraining influence therefore remain possible, and absolute performance may still be optimistic.

#### 6.2.3 External validity.

*Evaluation scope.* The resulting evidence supports transferability conclusions for six language ecosystems, mature logging-rich open-source repositories, seven LLMs, and one unified inference setup. Within the repository-snapshot benchmark, single-language leaderboards consistently identify the same top two models, while absolute performance levels, component-specific weaknesses, and fine-grained ordering below the top two vary across the evaluated language ecosystems. Generalization to private industrial codebases, lower-activity repositories, other language ecosystems, language-specialized adaptation strategies, or alternative prompting and retrieval configurations requires additional evaluation.

#### 6.2.4 Conclusion validity.

*Repeated-run and finite-sample uncertainty.* Temperature-0 inference does not eliminate service-side output variability, and repeated runs do not increase the number of benchmark cases. The revision-history branch is also much smaller and less balanced than the repository-snapshot branch. This imbalance can make language-level estimates and pairwise ranking correlations involving low-resource languages more sensitive to individual repositories and samples. Accordingly, RQ3 and RQ4 report the per-language revision-history composition, results from five repeated runs, standard deviations, and repository-clustered bootstrap intervals for revision-history ranking correlations and transformed-minus-original deltas.

*Composite-score sensitivity.* CCS provides a compact summary for ranking and trend analysis, but its gating structure and weight allocation encode design choices. Alternative composite formulations could therefore change some fine-grained outcomes, especially among closely matched middle-tier models and, under extreme PA-dominant settings, which language-specific subset yields the lowest cross-model mean. We report full component metrics alongside CCS, and RQ1.4 evaluates hierarchy-preserving alternative weightings. The main model-performance and ranking conclusions remain stable under these variants. CCS therefore serves as a summary of overall trends, while the component metrics provide the basis for interpreting specific generation behaviors.

*Metric conditioning.* Our component metrics follow a cascaded evaluation protocol: PA is computed on all samples, whereas FA, LA, message quality, and variable recovery are computed only on position-correct predictions. This design avoids scoring content at an incorrect logging site, but it also means that these metrics have language- and model-specific denominators and may be conditioned on different subsets of instance difficulty. FA differences across language-specific subsets therefore represent conditional lexical-anchor consistency once the logging site is correct, rather than unconditional framework selection over all inputs. Table 4 reports the position-correct subset sizes directly, while PA, CCS, and the component breakdowns jointly characterize conditional and end-to-end model performance.

### 6.3 Future Work

*Structure-aware and purpose-aware generation.* The difficulty of loop and nested-callable contexts suggests that future methods should model logging statement generation as local-purpose recovery rather than local code completion.

Incorporating richer structural signals, such as control-flow context, scope boundaries, and data dependencies, may help models infer which logging purpose is implied at a given site and which runtime state should be exposed.

*Ecosystem-aware retrieval and framework-grounded generation.* The increased framework-anchor dispersion in the revision-history setting motivates evaluating generation mechanisms that adapt to the target ecosystem and repository. This adaptation need not require a separate model for every programming language: a lightweight routing, prompting, or adapter layer could first identify the relevant logging framework, receiver, namespace, or macro family. Retrieval could then prioritize examples from the same language ecosystem and, where available, the same repository or logging framework, using imports, includes, logger declarations, nearby logging statements, and structural context as retrieval signals. A complementary verifier could resolve a generated anchor against in-scope symbols and known logging APIs, then rerank or repair inconsistent candidates. Future work should compare these alternatives against a shared multilingual baseline and report their effects on FA jointly with logging-site localization and the other logging-statement components, because improving framework alignment should not come at the cost of end-to-end generation quality.

*More realistic maintenance benchmarks.* Although the revision-history branch strengthens realism relative to repository-snapshot evaluation, it focuses on relatively clean local logging-statement-introducing edits. Future benchmarks could extend this setting to more complex maintenance scenarios, such as broader refactorings, multi-file changes, and logging updates intertwined with behavioral modification, to test whether multilingual trends remain stable under more demanding engineering conditions.

*Human-centered and tool-centered evaluation.* The practical value of automated logging statement generation depends on whether generated suggestions are useful to developers. An important next step is to complement benchmark-based evaluation with tool-centered studies, such as IDE-assisted logging support, code-review recommendation settings, post-edit acceptance analysis, and developer preference or usability studies, to connect multilingual benchmark performance to real-world maintenance benefit [35].

## 7 Conclusion

This paper presented MultiLogBench, a multilingual benchmark for automated logging statement generation across six language ecosystems. Using this benchmark, we examined whether conclusions about the task remain stable across language ecosystems and evaluation settings. The repository-snapshot results support three connected findings. First, the evaluated models' scores differ substantially across language-specific subsets, and the component breakdown shows that this conditional variation involves both logging-site localization and ecosystem-conditioned realization: position prediction has the largest average within-model component gap, while model outcomes exhibit distinct conditional content profiles. Second, among the primary structural buckets, loop and nested-callable contexts are the most difficult; the qualitative cases further suggest that context-localization difficulty contributes to this pattern. Third, the language-specific leaderboards have a mean pairwise Spearman correlation of 0.912 and preserve the same top two models across all six subsets, while middle- and lower-tier systems reorder. Automated logging is therefore language-sensitive in a claim-dependent way: absolute performance levels, component-specific weaknesses, and fine-grained ordering below the top two vary across language-specific subsets, while single-language leaderboards remain informative for identifying the leading models in the controlled repository-snapshot benchmark.

The revision-history evaluation shows which repository-snapshot findings persist under maintenance-derived conditions. In this setting, the six-language mean declines across all displayed metrics, the component profile shifts toward greater framework-anchor dispersion, and cross-language score dispersion persists, whereas the repository-snapshot top-two stability does not. Under the validated surface-distinct, target-preserving transformations, aggregate performance does not exhibit a broad same-direction collapse. Together, these findings show that multilingual evaluation is necessary for characterizing cross-language absolute performance, component-specific weaknesses, and fine-grained model selection, while maintenance-oriented validation determines which controlled-benchmark conclusions transfer to code evolution. Future work should target the observed failure sources: logging-site localization, framework-anchor and receiver alignment, structural localization, and local-purpose or target-state recovery.

## Acknowledgement

The work described in this paper was supported by two grants from the Research Grants Council of the Hong Kong Special Administrative Region, China: (1) No. CUHK 14209124 of the General Research Fund, and (2) No. SRFS2425-4S03 of the Senior Research Fellow Scheme.

## References

[1] Roozbeh Aghili, Heng Li, and Foutse Khomh. 2025. Protecting Privacy in Software Logs: What Should Be Anonymized? *Proc. ACM Softw. Eng.* 2, FSE, Article FSE060 (June 2025), 22 pages. https://doi.org/10.1145/3715779
[2] Anthropic. 2026. Introducing Claude Sonnet 4.6. Anthropic News. https://www.anthropic.com/news/claude-sonnet-4-6 Accessed 2026-04-10.
[3] Han Anu, Jie Chen, Wenchang Shi, Jianwei Hou, Bin Liang, and Bo Qin. 2019. An Approach to Recommendation of Verbosity Log Levels Based on Logging Intention. In *2019 IEEE International Conference on Software Maintenance and Evolution (ICSME)*. IEEE, 125–134. https://doi.org/10.1109/icsme.2019.00022
[4] Max Brunsfeld and tree-sitter contributors. 2026. tree-sitter: An Incremental Parsing System for Programming Tools. https://github.com/tree-sitter/tree-sitter. GitHub repository. Accessed April 17, 2026.
[5] Jeanderson Candido, Jan Haesen, Mauricio Aniche, and Arie van Deursen. 2021. An Exploratory Study of Log Placement Recommendation in an Enterprise System. In *2021 IEEE/ACM 18th International Conference on Mining Software Repositories (MSR)*. IEEE, 143–154. https://doi.org/10.1109/msr52588.2021.00027
[6] Boyuan Chen and Zhen Ming Jiang. 2017. Characterizing and detecting anti-patterns in the logging code. In *2017 IEEE/ACM 39th International Conference on Software Engineering (ICSE)*. IEEE, 71–81. https://doi.org/10.1109/icse.2017.15
[7] Boyuan Chen and Zhen Ming Jiang. 2017. Characterizing logging practices in Java-based open source software projects - a replication study in Apache Software Foundation. *Empirical Software Engineering* 22, 1 (2017), 330–374. https://doi.org/10.1007/s10664-016-9429-5
[8] Boyuan Chen and Zhen Ming Jiang. 2019. Extracting and studying the Logging-Code-Issue-Introducing changes in Java-based large-scale open source software systems. *Empirical Software Engineering* 24, 4 (2019), 2285–2322. https://doi.org/10.1007/s10664-019-09690-0
[9] Boyuan Chen and Zhen Ming Jiang. 2020. Studying the use of Java logging utilities in the wild. In *Proceedings of the ACM/IEEE 42nd International Conference on Software Engineering*. ACM, 397–408. https://doi.org/10.1145/3377811.3380408
[10] Ozren Dabic, Emad Aghajani, and Gabriele Bavota. 2021. Sampling projects in github for MSR studies. In *2021 IEEE/ACM 18th International Conference on Mining Software Repositories (MSR)*. IEEE, 560–564. https://doi.org/10.1109/MSR52588.2021.00074
[11] DeepSeek-AI. 2025. DeepSeek-V3.2: Pushing the Frontier of Open Large Language Models. https://doi.org/10.48550/arXiv.2512.02556 arXiv:2512.02556 [cs.CL]
[12] Zishuo Ding, Heng Li, and Weiyi Shang. 2022. LoGenText: Automatically Generating Logging Texts Using Neural Machine Translation. In *2022 IEEE International Conference on Software Analysis, Evolution and Reengineering (SANER)*. IEEE, 349–360. https://doi.org/10.1109/saner53432.2022.00051
[13] Zishuo Ding, Yiming Tang, Xiaoyu Cheng, Heng Li, and Weiyi Shang. 2023. LoGenText-Plus: Improving Neural Machine Translation Based Logging Texts Generation with Syntactic Templates. *ACM Transactions on Software Engineering and Methodology* 33, 2 (2023), 1–45. https://doi.org/10.1145/3624740
[14] Zishuo Ding, Yiming Tang, Yang Li, Heng Li, and Weiyi Shang. 2023. On the temporal relations between logging and code. In *2023 IEEE/ACM 45th International Conference on Software Engineering (ICSE)*. IEEE, 843–854. https://doi.org/10.1109/ICSE48619.2023.00079
[15] Shengcheng Duan, Yihua Xu, Sheng Zhang, Shen Wang, and Yue Duan. 2025. PDLogger: Automated Logging Framework for Practical Software Development. *arXiv preprint arXiv:2507.19951* (2025). https://doi.org/10.48550/arXiv.2507.19951
[16] Patrick Loic Foalem, Foutse Khomh, and Heng Li. 2024. Studying logging practice in machine learning-based applications. *Information and Software Technology* 170 (2024), 107450. https://doi.org/10.1016/j.infsof.2024.107450
[17] Qiang Fu, Jieming Zhu, Wenlu Hu, Jian-Guang Lou, Rui Ding, Qingwei Lin, Dongmei Zhang, and Tao Xie. 2014. Where do developers log? an empirical study on logging practices in industry. In *Companion Proceedings of the 36th International Conference on Software Engineering*. ACM, 24–33. https://doi.org/10.1145/2591062.2591175

[18] Sina Gholamian and Paul A. S. Ward. 2020. Logging statements' prediction based on source code clones. In *Proceedings of the 35th Annual ACM Symposium on Applied Computing (SAC '20)*. ACM, 82–91. https://doi.org/10.1145/3341105.3373845

[19] GitHub. 2025. Octoverse: A New Developer Joins GitHub Every Second as AI Leads TypeScript to #1. https://github.blog/news-insights/octoverse/octoverse-a-new-developer-joins-github-every-second-as-ai-leads-typescript-to-1/. GitHub Blog. Published October 28, 2025. Accessed April 3, 2026.

[20] GLM-5-Team, Aohan Zeng, et al. 2026. GLM-5: from Vibe Coding to Agentic Engineering. https://doi.org/10.48550/arXiv.2602.15763 arXiv:2602.15763 [cs.LG]

[21] Wenwei Gu, Renyi Zhong, Guangba Yu, Xinying Sun, Jinyang Liu, Yintong Huo, Zhuangbin Chen, Jianping Zhang, Jiazhen Gu, Yongqiang Yang, and Michael R. Lyu. 2025. KPIRoot+: An Efficient Integrated Framework for Anomaly Detection and Root Cause Analysis in Large-Scale Cloud Systems. *Empirical Software Engineering* 31, 2 (2025), 28. https://doi.org/10.1007/s10664-025-10769-0

[22] Mehran Hassani, Weiyi Shang, Emad Shihab, and Nikolaos Tsantalis. 2018. Studying and detecting log-related issues. *Empirical Software Engineering* 23, 6 (2018), 3248–3280. https://doi.org/10.1007/s10664-018-9603-z

[23] Shilin He, Pinjia He, Zhuangbin Chen, Tianyi Yang, Yuxin Su, and Michael R Lyu. 2021. A survey on automated log analysis for reliability engineering. *ACM computing surveys (CSUR)* 54, 6 (2021), 1–37. https://doi.org/10.1145/3460345

[24] Zhihan Jiang, Junjie Huang, Guangba Yu, Zhuangbin Chen, Yichen Li, Renyi Zhong, Cong Feng, Yongqiang Yang, Zengyin Yang, and Michael Lyu. 2025. L4: Diagnosing large-scale llm training failures via automated log analysis. In *Proceedings of the 33rd ACM International Conference on the Foundations of Software Engineering*. 51–63. https://doi.org/10.1145/3696630.3728531

[25] Suhas Kabinna, Cor-Paul Bezemer, Weiyi Shang, and Ahmed E. Hassan. 2016. Logging library migrations: a case study for the Apache Software Foundation projects. In *Proceedings of the 13th International Conference on Mining Software Repositories*. ACM, 154–164. https://doi.org/10.1145/2901739.2901769

[26] Suhas Kabinna, Cor-Paul Bezemer, Weiyi Shang, and Ahmed E. Hassan. 2018. Examining the stability of logging statements. *Empirical Software Engineering* 23, 1 (2018), 290–333. https://doi.org/10.1007/s10664-017-9518-0

[27] Taeyoung Kim, Suntae Kim, Sooyong Park, and YoungBeom Park. 2019. Automatic recommendation to appropriate log levels. *Software: Practice and Experience* 50, 3 (2019), 189–209. https://doi.org/10.1002/spe.2771

[28] Kimi Team, Tongtong Bai, et al. 2026. Kimi K2.5: Visual Agentic Intelligence. https://doi.org/10.48550/arXiv.2602.02276 arXiv:2602.02276 [cs.CL]

[29] Heng Li, Weiyi Shang, Bram Adams, Mohammed Sayagh, and Ahmed E. Hassan. 2021. A Qualitative Study of the Benefits and Costs of Logging From Developers' Perspectives. *IEEE Transactions on Software Engineering* 47, 12 (2021), 2858–2873. https://doi.org/10.1109/tse.2020.2970422

[30] Heng Li, Weiyi Shang, and Ahmed E. Hassan. 2016. Which log level should developers choose for a new logging statement? *Empirical Software Engineering* 22, 4 (2016), 1684–1716. https://doi.org/10.1007/s10664-016-9456-2

[31] Heng Li, Haoxiang Zhang, Shaowei Wang, and Ahmed E. Hassan. 2022. Studying the Practices of Logging Exception Stack Traces in Open-Source Software Projects. *IEEE Transactions on Software Engineering* 48, 12 (2022), 4907–4924. https://doi.org/10.1109/tse.2021.3129688

[32] Min Li, Gou Tan, Pengfei Chen, and Chuanfu Zhang. 2026. LogGen: Integrating traditional model and LLM with code analysis for precise log generation. *Journal of Systems and Software* 236 (2026), 112816. https://doi.org/10.1016/j.jss.2026.112816

[33] Yichen Li, Yintong Huo, Zhihan Jiang, Renyi Zhong, Pinjia He, Yuxin Su, Lionel C. Briand, and Michael R. Lyu. 2024. Exploring the Effectiveness of LLMs in Automated Logging Statement Generation: An Empirical Study. *IEEE Transactions on Software Engineering* 50, 12 (2024), 3188–3207. https://doi.org/10.1109/tse.2024.3475375

[34] Yichen Li, Yintong Huo, Renyi Zhong, Zhihan Jiang, Jinyang Liu, Junjie Huang, Jiazhen Gu, Pinjia He, and Michael R. Lyu. 2024. Go Static: Contextualized Logging Statement Generation. *Proceedings of the ACM on Software Engineering* 1, FSE (2024), 609–630. https://doi.org/10.1145/3643754

[35] Yichen Li, Jinyang Liu, Junsong Pu, Zhihan Jiang, Zhuangbin Chen, Xiao He, Tieying Zhang, Jianjun Chen, Yi Li, Rui Shi, and Michael R. Lyu. 2025. Automated Proactive Logging Quality Improvement for Large-Scale Codebases. In *2025 40th IEEE/ACM International Conference on Automated Software Engineering (ASE)*. 3426–3437. https://doi.org/10.1109/ASE63991.2025.00283

[36] Zhenhao Li, An Ran Chen, Xing Hu, Xin Xia, Tse-Hsun Chen, and Weiyi Shang. 2023. Are they all good? studying practitioners' expectations on the readability of log messages. In *2023 38th IEEE/ACM International Conference on Automated Software Engineering (ASE)*. IEEE, 129–140. https://doi.org/10.1109/ASE56229.2023.00136

[37] Zhenhao Li, Tse-Hsun Chen, Jinqiu Yang, and Weiyi Shang. 2019. DLFinder: Characterizing and Detecting Duplicate Logging Code Smells. In *2019 IEEE/ACM 41st International Conference on Software Engineering (ICSE)*. IEEE, 152–163. https://doi.org/10.1109/icse.2019.00032

[38] Zhenhao Li, Tse-Hsun Chen, Jinqiu Yang, and Weiyi Shang. 2021. Studying duplicate logging statements and their relationships with code clones. *IEEE Transactions on Software Engineering* 48, 7 (2021), 2476–2494. https://doi.org/10.1109/TSE.2021.3060918

[39] Zhenhao Li, Tse-Hsun (Peter) Chen, and Weiyi Shang. 2020. Where shall we log?: studying and suggesting logging locations in code blocks. In *Proceedings of the 35th IEEE/ACM International Conference on Automated Software Engineering (ASE)*. ACM, Virtual Event Australia. https://doi.org/10.1145/3324884.3416636

[40] Zhenhao Li, Heng Li, Tse-Hsun Chen, and Weiyi Shang. 2021. Deeplv: Suggesting log levels using ordinal based neural networks. In *2021 IEEE/ACM 43rd International Conference on Software Engineering (ICSE)*. IEEE, 1461–1472. https://doi.org/10.1109/ICSE43902.2021.00131

[41] Chin-Yew Lin. 2004. ROUGE: A Package for Automatic Evaluation of Summaries. In *Text Summarization Branches Out*. Association for Computational Linguistics, Barcelona, Spain, 74–81. https://aclanthology.org/W04-1013/

[42] Jiahao Liu, Jun Zeng, Xiang Wang, Kaihang Ji, and Zhenkai Liang. 2022. TeLL: log level suggestions via modeling multi-level code block information. In *Proceedings of the 31st ACM SIGSOFT International Symposium on Software Testing and Analysis (ISSTA '22)*. ACM, 27–38. https://doi.org/10.1145/3533767.3534379

[43] Zhongxin Liu, Xin Xia, David Lo, Zhenchang Xing, Ahmed E. Hassan, and Shanping Li. 2019. Which Variables Should I Log? *IEEE Transactions on Software Engineering (TSE)* (2019), 1–1. https://doi.org/10.1109/TSE.2019.2941943

[44] Antonio Mastropaolo, Valentina Ferrari, Luca Pascarella, and Gabriele Bavota. 2024. Log statements generation via deep learning: Widening the support provided to developers. *Journal of Systems and Software* 210 (2024), 111947. https://doi.org/10.1016/j.jss.2023.111947

[45] Antonio Mastropaolo, Luca Pascarella, and Gabriele Bavota. 2022. Using deep learning to generate complete log statements. In *Proceedings of the 44th International Conference on Software Engineering* (Pittsburgh, Pennsylvania) *(ICSE '22)*. Association for Computing Machinery, New York, NY, USA, 2279–2290. https://doi.org/10.1145/3510003.3511561

[46] Alexandre Matton, Tom Sherborne, Dennis Aumiller, Elena Tommasone, Milad Alizadeh, Jingyi He, Raymond Ma, Maxime Voisin, Ellen Gilsenan-McMahon, and Matthias Gallé. 2024. On Leakage of Code Generation Evaluation Datasets. In *Findings of the Association for Computational Linguistics: EMNLP 2024*. Association for Computational Linguistics, 13215–13223. https://doi.org/10.18653/v1/2024.findings-emnlp.772

[47] OpenAI. 2025. Advanced usage: Reproducible outputs. https://platform.openai.com/docs/advanced-usage/reproducible-outputs. Accessed: 2025-11-30.

[48] OpenAI. 2026. Introducing GPT-5.4. OpenAI Product Release. https://openai.com/index/introducing-gpt-5-4/ Accessed 2026-04-10.

[49] Kishore Papineni, Salim Roukos, Todd Ward, and Wei-Jing Zhu. 2002. Bleu: a method for automatic evaluation of machine translation. In *Proceedings of the 40th annual meeting of the Association for Computational Linguistics (ACL)*. 311–318. https://doi.org/10.3115/1073083.1073135

[50] Qwen Team. 2025. Qwen3 Technical Report. https://doi.org/10.48550/arXiv.2505.09388 arXiv:2505.09388 [cs.CL]

[51] Martin Riddell, Ansong Ni, and Arman Cohan. 2024. Quantifying Contamination in Evaluating Code Generation Capabilities of Language Models. In *Proceedings of the 62nd Annual Meeting of the Association for Computational Linguistics (Volume 1: Long Papers)*. Association for Computational Linguistics, 14116–14137. https://doi.org/10.18653/v1/2024.acl-long.761

[52] Stephen Robertson and Hugo Zaragoza. 2009. *The probabilistic relevance framework: BM25 and beyond*. Vol. 4. Now Publishers Inc. https://doi.org/10.1561/1500000019

[53] Guoping Rong, Shenghui Gu, He Zhang, Dong Shao, and Wanggen Liu. 2018. How Is Logging Practice Implemented in Open Source Software Projects? A Preliminary Exploration. In *2018 25th Australasian Software Engineering Conference (ASWEC)*. IEEE, 171–180. https://doi.org/10.1109/aswec.2018.00031

[54] Guoping Rong, Yangchen Xu, Shenghui Gu, He Zhang, and Dong Shao. 2020. Can You Capture Information As You Intend To? A Case Study on Logging Practice in Industry. In *2020 IEEE International Conference on Software Maintenance and Evolution (ICSME)*. IEEE, 12–22. https://doi.org/10.1109/icsme46990.2020.00012

[55] June Sallou, Thomas Durieux, and Annibale Panichella. 2024. Breaking the silence: the threats of using llms in software engineering. In *Proceedings of the 2024 ACM/IEEE 44th International Conference on Software Engineering: New Ideas and Emerging Results*. 102–106. https://doi.org/10.1145/3639476.3639764

[56] Weiyi Shang, Meiyappan Nagappan, and Ahmed E. Hassan. 2015. Studying the relationship between logging characteristics and the code quality of platform software. *Empirical Software Engineering* 20, 1 (2015), 1–27. https://doi.org/10.1007/s10664-013-9274-8

[57] Weiyi Shang, Meiyappan Nagappan, Ahmed E. Hassan, and Zhen Ming Jiang. 2014. Understanding Log Lines Using Development Knowledge. In *2014 IEEE International Conference on Software Maintenance and Evolution*. IEEE, 21–30. https://doi.org/10.1109/icsme.2014.24

[58] Stack Overflow. 2024. 2024 Stack Overflow Developer Survey: Technology. https://survey.stackoverflow.co/2024/technology. Accessed April 3, 2026.

[59] Boyin Tan, Junjielong Xu, Zhouruixing Zhu, and Pinjia He. 2025. AL-Bench: A Benchmark for Automatic Logging. https://doi.org/10.48550/ARXIV.2502.03160

[60] Yiming Tang, Allan Spektor, Raffi Khatchadourian, and Mehdi Bagherzadeh. 2022. Automated evolution of feature logging statement levels using git histories and degree of interest. *Science of Computer Programming* 214 (2022), 102724. https://doi.org/10.1016/j.scico.2021.102724

[61] The Gemini Team. 2026. Gemini 3.1 Pro: A Smarter Model for Your Most Complex Tasks. Google Blog. https://blog.google/innovation-and-ai/models-and-research/gemini-models/gemini-3-1-pro/ Accessed 2026-04-10.

[62] Xin Wang, Yang Feng, Jiaoxiao Qian, Yang Zhang, Zhenhao Li, and Zishuo Ding. 2026. Logging Like Humans for LLMs: Rethinking Logging via Execution and Runtime Feedback. https://doi.org/10.48550/arXiv.2603.29122 arXiv:2603.29122 [cs.SE]

[63] Xin Wang, Zhenhao Li, and Zishuo Ding. 2025. Defects4Log: Benchmarking LLMs for Logging Code Defect Detection and Reasoning. *arXiv preprint arXiv:2508.11305* (2025). https://doi.org/10.48550/arXiv.2508.11305

[64] Yi Wu, Nan Jiang, Hung Viet Pham, Thibaud Lutellier, Jordan Davis, Lin Tan, Petr Babkin, and Sameena Shah. 2023. How effective are neural networks for fixing security vulnerabilities. In *Proceedings of the 32nd ACM SIGSOFT International Symposium on Software Testing and Analysis*. 1282–1294. https://doi.org/10.1145/3597926.3598135

[65] Xiaoyuan Xie, Zhipeng Cai, Songqiang Chen, and Jifeng Xuan. 2024. FastLog: An End-to-End Method to Efficiently Generate and Insert Logging Statements. In *Proceedings of the 33rd ACM SIGSOFT International Symposium on Software Testing and Analysis (ISSTA '24)*. ACM, 26–37. https://doi.org/10.1145/3650212.3652107

[66] Junjielong Xu, Ziang Cui, Yuan Zhao, Xu Zhang, Shilin He, Pinjia He, Liqun Li, Yu Kang, Qingwei Lin, Yingnong Dang, Saravan Rajmohan, and Dongmei Zhang. 2024. UniLog: Automatic Logging via LLM and In-Context Learning. In *Proceedings of the IEEE/ACM 46th International Conference*

*on Software Engineering (ICSE '24)*. ACM, 1–12. https://doi.org/10.1145/3597503.3623326

[67] Yi Zeng, Jinfu Chen, Weiyi Shang, and Tse-Hsun Chen. 2019. Studying the characteristics of logging practices in mobile apps: a case study on F-Droid. *Empirical Software Engineering* 24, 6 (2019), 3394–3434. https://doi.org/10.1007/s10664-019-09687-9

[68] Haonan Zhang, Yiming Tang, Maxime Lamothe, Heng Li, and Weiyi Shang. 2022. Studying Logging Practice in Test Code. *Empirical Software Engineering* 27, 4 (2022), 83. https://doi.org/10.1007/s10664-022-10139-0

[69] Xu Zhao, Kirk Rodrigues, Yu Luo, Michael Stumm, Ding Yuan, and Yuanyuan Zhou. 2017. Log20: Fully Automated Optimal Placement of Log Printing Statements under Specified Overhead Threshold. In *Proceedings of the 26th Symposium on Operating Systems Principles (SOSP)*. ACM, Shanghai China, 565–581. https://doi.org/10.1145/3132747.3132778

[70] Chen Zhi, Jianwei Yin, Shuiguang Deng, Maoxin Ye, Min Fu, and Tao Xie. 2019. An Exploratory Study of Logging Configuration Practice in Java. In *2019 IEEE International Conference on Software Maintenance and Evolution (ICSME)*. IEEE, 459–469. https://doi.org/10.1109/icsme.2019.00079

[71] Chen Zhi, Jianwei Yin, Junxiao Han, and Shuiguang Deng. 2020. A Preliminary Study on Sensitive Information Exposure Through Logging. In *2020 27th Asia-Pacific Software Engineering Conference (APSEC)*. IEEE, 470–474. https://doi.org/10.1109/apsec51365.2020.00058

[72] Renyi Zhong, Yintong Huo, Wenwei Gu, Yichen Li, and Michael R. Lyu. 2025. End-to-End Automated Logging via Multi-Agent Framework. https://doi.org/10.48550/arXiv.2511.18528 arXiv:2511.18528 [cs.SE]

[73] Renyi Zhong, Yichen Li, Jinxi Kuang, Wenwei Gu, Yintong Huo, and Michael R. Lyu. 2025. LogUpdater: Automated Detection and Repair of Specific Defects in Logging Statements. *ACM Trans. Softw. Eng. Methodol.* 35, 1, Article 16 (Dec. 2025), 31 pages. https://doi.org/10.1145/3731754

[74] Renyi Zhong, Yichen Li, Guangba Yu, Wenwei Gu, Jinxi Kuang, Yintong Huo, and Michael R. Lyu. 2025. Larger Is Not Always Better: Exploring Small Open-source Language Models in Logging Statement Generation. *ACM Transactions on Software Engineering and Methodology* (2025). https://doi.org/10.1145/3773287

[75] Rui Zhou, Mohammad Hamdaqa, Haipeng Cai, and Abdelwahab Hamou-Lhadj. 2020. MobiLogLeak: A Preliminary Study on Data Leakage Caused by Poor Logging Practices. In *2020 IEEE International Conference on Software Analysis, Evolution and Reengineering (SANER)*. IEEE, 577–581. https://doi.org/10.1109/saner48275.2020.9054831

[76] Jieming Zhu, Pinjia He, Qiang Fu, Hongyu Zhang, Michael R. Lyu, and Dongmei Zhang. 2015. Learning to Log: Helping Developers Make Informed Logging Decisions. In *2015 IEEE/ACM 37th IEEE International Conference on Software Engineering*. IEEE, 415–425. https://doi.org/10.1109/icse.2015.60